\documentclass[10pt, a4paper,twoside]{IEEEtran}

\usepackage{cite}
\usepackage{amsmath,amssymb,amsfonts}
\usepackage{algorithmic}
\usepackage{graphicx}
\usepackage{subfig}
\usepackage{algorithm,algorithmic}
\usepackage{hyperref}
\usepackage{textcomp}

\def\BibTeX{{\rm B\kern-.05em{\sc i\kern-.025em b}\kern-.08em
    T\kern-.1667em\lower.7ex\hbox{E}\kern-.125emX}}

\newtheorem{theorem}{Theorem}[section]

\newtheorem{corollary}{Corollary}[section]

\newtheorem{definition}{Definition}[section]
\newtheorem{example}{Example}[section]
\newtheorem{exercise}{Exercise}[section]
\newtheorem{lemma}{Lemma}[section]

\newtheorem{problem}{Problem}[section]
\newtheorem{proposition}{Proposition}[section]
\newtheorem{remark}{Remark}[section]
\newtheorem{assumption}{Assumption}[section]
\newcommand{\bthm}{\begin{theorem}}
\newcommand{\ethm}{\end{theorem}}
\newcommand{\blem}{\begin{lemma}}
\newcommand{\elem}{\end{lemma}}
\newcommand{\bex}{\begin{example}}
\newcommand{\eex}{\end{example}}
\newcommand{\beg}{\begin{exercise}}
\newcommand{\eeg}{\end{exercise}}
\newcommand{\bprop}{\begin{proposition}}
\newcommand{\eprop}{\end{proposition}}
\newcommand{\bplm}{\begin{problem}}
\newcommand{\eplm}{\end{problem}}
\newcommand{\bmrk}{\begin{remark}}
\newcommand{\emrk}{\end{remark}}
\newcommand{\bdfn}{\begin{definition}}
\newcommand{\edfn}{\end{definition}}
\newcommand{\bcor}{\begin{corollary}}
\newcommand{\ecor}{\end{corollary}}

\newcommand{\beq}{\begin{equation}}
\newcommand{\eeq}{\end{equation}}
\newcommand{\beqm}{\begin{equation*}}
\newcommand{\eeqm}{\end{equation*}}
\newcommand{\beqn}{\begin{eqnarray}}
\newcommand{\eeqn}{\end{eqnarray}}
\newcommand{\beqnm}{\begin{eqnarray*}}
\newcommand{\eeqnm}{\end{eqnarray*}}
\newcommand{\bea}{\begin{aligned}}
\newcommand{\eea}{\end{aligned}}
\newcommand{\beam}{\begin{aligned*}}
\newcommand{\eeam}{\end{aligned*}}
\newcommand{\bs}{\begin{subequations}}
\newcommand{\es}{\end{subequations}}

\newcommand{\bei}{\begin{itemize}}
\newcommand{\eei}{\end{itemize}}
\newcommand{\bed}{\begin{description}}
\newcommand{\eed}{\end{description}}
\newcommand{\bee}{\begin{enumerate}}
\newcommand{\eee}{\end{enumerate}}
\newcommand{\bey}{\begin{array}}
\newcommand{\eey}{\end{array}}
\newcommand{\bec}{\begin{center}}
\newcommand{\eec}{\end{center}}
\newcommand{\la}{\label}

\newcommand{\mbb}{\mathbb}
\newcommand{\mbf}{\mathbf}

\def\l{\langle}
\def\r{\rangle}

\begin{document}
\title{Linear quantum systems: poles, zeros, invertibility and sensitivity}
\author{Zhiyuan Dong, Guofeng Zhang, \IEEEmembership{Senior member, IEEE},  Heung-wing Joseph Lee, \\
and Ian R. Petersen, \IEEEmembership{Fellow, IEEE}
\thanks{Corresponding author: Guofeng Zhang.}
\thanks{Zhiyuan Dong is with School of Science, Harbin Institute of Technology, Shenzhen, China (e-mail: dongzhiyuan@hit.edu.cn).}
\thanks{Guofeng Zhang is with Department of Applied Mathematics, The Hong Kong Polytechnic University, Hong Kong, China and also with Shenzhen Research Institute, The Hong Kong Polytechnic University, Shenzhen, China (e-mail: guofeng.zhang@polyu.edu.hk).}
\thanks{Heung-wing Joseph Lee is with Department of Applied Mathematics, The Hong Kong Polytechnic University, Hong Kong, China (e-mail: joseph.lee@polyu.edu.hk).}
\thanks{Ian R. Petersen is with School of Engineering, The Australian National University, Canberra ACT 2601, Australia (e-mail: i.r.petersen@gmail.com).}
}

\maketitle

\begin{abstract}
The non-commutative nature of quantum mechanics imposes fundamental constraints on system dynamics, which, in the linear realm, are manifested through the physical realizability conditions on system matrices. These restrictions give system matrices a unique structure. This paper aims to study this structure by investigating the zeros and poles of linear quantum systems. Firstly, it is shown that $-s_0$ is a transmission zero if and only if $s_0$ is a pole of the transfer function, and $-s_0$  is an invariant zero if and only if  $s_0$  is an eigenvalue of the  $A$-matrix, of a linear quantum system. Moreover,  $s_0$ is an output-decoupling zero if and only if $-s_0$ is an input-decoupling zero. Secondly, based on these pole-zero correspondences and inspired by a recent work on the stable inversion of classical linear systems \cite{DD2023}, we show that a linear quantum system must be Hurwitz unstable if it is strongly asymptotically left invertible. Two types of stable input observers are constructed for unstable linear quantum systems. Finally, the sensitivity of a coherent feedback network is investigated; in particular, the fundamental tradeoff between ideal squeezing and system robustness is studied on the basis of system sensitivity analysis.
\end{abstract}


\section{Introduction}

In recent decades, significant advancements have been made in both theoretical understanding and experimental demonstrations of quantum control. Quantum control plays a key  role in a variety of quantum information technologies, such as quantum communication, quantum computation, quantum cryptography, quantum ultra-precision metrology, and nano-electronics \cite{NC10,WM10,NY17,DP22,BQND24}. Analogous to classical control systems theory, linear quantum systems hold great importance in the field of quantum control. Linear quantum systems are mathematical models that describe the behavior of quantum harmonic oscillators. In this context, ``linear'' refers to the linearity of the Heisenberg equations of motion for quadrature operators in the quantum system. This linearity often leads to simplifications that facilitate the analysis and control of these systems \cite{NY14,ZD22,BQD24}. Consequently, linear quantum systems can be effectively studied using powerful mathematical techniques derived from linear systems theory. A wide range of quantum-mechanical systems can be nicely modeled as linear quantum systems; for instance, quantum optical systems  \cite{WM94,GZ00,WM08,Mabuchi08,WM10,ZJ12,P14,CKS17,NY17,PJU+20,BNC+21},
circuit quantum electro-dynamical (circuit QED) systems \cite{MJP+11,BLS+11,KAK13,BGGW21}, cavity QED systems \cite{DJ99,SDZ+11,ASD+12}, quantum opto-mechanical systems  \cite{TC10,MHP+11,HM12,DFK+12,MCP+12,NY13,NY14,AKM14,ODP+16,TBCGKP2020,KPS+21,LOW+21,SQH+23},
atomic ensembles \cite{SvHM04,NJP09,NY13,ANP+17,TBCGKP2020}, and quantum memories \cite{XDL07,HRG+09,HCH+13,YJ14,NG15}.

Although formally similar to linear classical systems, linear quantum systems possess a unique structure due to the non-commutative nature of quantum mechanics. The system matrices of a linear quantum system have a very special structure, which endows linear quantum systems with distinct properties compared to their classical counterparts. For instances,   for
linear quantum passive systems, controllability is equivalent to observability and they imply Hurwitz stability \cite[Lem. 3.1]{GY15}. In fact, later it is proved in \cite[Lem. 2 and  Thm. 2]{GZ15} that  controllability, observability and Hurwitz stability are all equivalent for linear quantum passive systems.  For general linear quantum systems (not necessarily passive),  controllability and observability are equivalent \cite[Prop. 1]{GZ15}. However, for linear quantum systems stabilizability and detectability are  not equivalent, as demonstrated by Example  \ref{eq:Example_h} in this paper, where the system is detectable but not  stabilizable.  Moreover, if a linear quantum system is Hurwitz stable, then it is both controllable and observable \cite[Thm. 3.1]{ZPL20}. Finally, it is shown in \cite{ZGPG18} that the controllable and unobservable (``$c\bar{o}$'') subsystem and the uncontrollable and observable (``$\bar{c}o$'') subsystem coexist or vanish simultaneously.

In systems and control theory, zeros and poles are important concepts  which play a significant role in the dynamics and controller design of linear dynamical systems. There are several definitions of system zeros, including decoupling zero \cite{HJ84}, blocking zero \cite{FB77,FL85,BD85,Patel86,CHEN1992}, transmission zero and invariant zero \cite{kailath1980,DCC1982,CD1991,ZDG96,SP05}. Roughly speaking, a transmission zero represents the complex frequency at which the system's output is zero in response to inputs in a certain direction (called zero input direction). Non-minimum phase (namely zeros in the  right-half plane) and unstable poles often pose fundamental performance limitations in control systems design  \cite{bode1945network,SP05,SBG12,CI19,KS19}. System zeros also play a key role in the system inversion theory, which aims to estimate/reconstruct the input of a system based on its output. Simply speaking, if a classical linear system is minimum phase, then under mild conditions the system has a stable inverse so that the input can be estimated/reconstructed from its output \cite{Moylan77,Zames81,Hautus83,HP98,Sogo10,CP18,RG19,DD2023}.      Recently, strong left-invertibility of linear systems is characterized by means of invariant zeros \cite{DD2023,MIB23}, and  constructive procedures for input reconstruction are proposed.

It is very plausible to speculate that poles and zeros are important also in linear quantum systems theory \cite{YJ14,ZD22} and are closely related to fundamental performance limitations of linear quantum systems. It fact, it has been shown in  \cite{Yan07} that  these concepts are fundamental to understanding the behavior of linear quantum systems. The purpose of this paper is to delve into the intricacies of invariant zeros, transmission zeros, invertibility, and sensitivity of linear quantum systems, elucidating their roles, relationships, and implications in quantum control theory. Firstly, we prove that $-s_0$ is a transmission zero if and only if $s_0$ is a pole of the transfer matrix of a linear quantum system, and in analog $-s_0$ is an invariant zero if and only if $s_0$ is an eigenvalue of the $A$-matrix  of a linear quantum system. Moreover, $-s_0$ is an output-decoupling zero if and only if $s_0$ is an input-decoupling zero. Secondly, based on these pole-zero correspondences, we show that a linear quantum system must be Hurwitz unstable if it is asymptotically strongly left invertible.  Moreover, two different types of stable input observers are constructed for unstable linear quantum systems. The first type of stable input observers  are classical, but they different from the recently constructed one \cite{DD2023}. The second type is of  quantum-mechanical nature. Finally, the relation between quantum squeezing and sensitivity of a coherent feedback network is investigated.

Some of the results of this article were presented in the conference paper \cite{DZL24}. The following are the most notable comparisons.

\begin{itemize}
\item
In Section \ref{Preli}, Example \ref{eq:Example_h} is a new result. Also, Definitions \ref{def:stab_ctrb_obsv} - \ref{def:pole} are given to streamline the discussions in the sequel.
\item
In Section \ref{ZPLQS}, Corollary \ref{prop_G and inv G_real}, Proposition \ref{Feb3-3} are Propositions 3.1 and 3.2 in \cite{DZL24}, respectively. However, Propositions \ref{odzero}, \ref{prop_G and inv G} and \ref{prop3.4}, Corollary \ref{cor3.1}, Theorems \ref{cor:s_-s} and \ref{cor:invariant_0} are new results. In addition, we have refined the statement of Theorem 3.1 in \cite{DZL24}, which becomes Theorem \ref{cor:union_eig} in this paper to enhance mathematical rigor. Moreover, Examples \ref{exam3.2}, \ref{ex:3.2}, and \ref{ex:bar_co} are new results.
\item
In Section \ref{IZLI}, Corollary \ref{corollary4.4} is Remark 4.3 in \cite{DZL24}. However, Theorem 4.1 in \cite{DZL24} has been restated as Theorem \ref{Thm6.1} with enhanced precision to explicitly quantify the necessary and sufficient condition.
Corollaries \ref{corollary4.1}, \ref{cor:A_co}, \ref{cor:minimal}, and Example \ref{example_4.1} are new results. Moreover, the entire Subsections \ref{eq:subsec:observer} and \ref{eq:subsec:Strong}  are new results.
\item
In Section \ref{TSQS}, the quantum ideal squeezing through a coherent feedback network, carried out in Subsections \ref{subsec:the network} and \ref{subsec:squeezing}, is a much more comprehensive extension of those in Section V of \cite{DZL24}. In particular, Theorems \ref{thm:blocking_zero_squeezing}  and \ref{thm:squeezing_alpha} are new results. Moreover, the entire Subsection \ref{subsec:squeezing_robustness}  is new, in which Theorem \ref{thm:S+T} is given and  a simulation study illustrating pole-zero duality is also conducted in Example \ref{Ex.5.4} and Fig. \ref{fig:total_figure}.
\item
Section \ref{conclu} expands the conclusion in \cite{DZL24} by giving a more detailed discussion the pole-zero duality and its applications on input reconstruction and sensitivity analysis for linear quantum systems.
\end{itemize}

The rest of this paper is organized as follows. Linear quantum systems and their poles and zeros are briefly introduced in Section \ref{Preli}. The special structural properties of linear quantum systems in terms of their zeros and poles are investigated in Section \ref{ZPLQS}. Section \ref{IZLI} discusses left invertibility of linear quantum systems by means of invariant zeros. Tradeoffs between quantum squeezing and system robustness  of  a coherent feedback network is studied in Section \ref{TSQS}. Section \ref{conclu} concludes this paper.

\emph{Notation.}
Let $\imath=\sqrt{-1}$ denote the imaginary unit. $\mathbb{R}$ is the field of real numbers,  $\mathbb{C}$ is the field of complex numbers, and $\mathbb{Z}^+$ is the set of positive integers. The set of eigenvalues of a matrix $A$ is denoted $\lambda(A)$. Given two matrices $A$ and $B$, denote $\lambda(A)/\lambda(B) = \{\lambda:  \lambda \in \lambda(A), \lambda \not\in  \lambda(B)  \}$.  For a column vector of complex numbers or operators $X=[x_1,\ldots,x_n]^\top$, the complex conjugate of $X$ or its adjoint operator is denoted by $X^\#=[x_1^\ast,\ldots,x_n^\ast]^\top$. Denote  $X^\dagger=(X^\#)^\top$ and  $\breve{X}=[
 X^\top \ X^\dagger ]^\top$. For two matrices $U,V\in\mathbb{C}^{k\times r}$, define the doubled-up matrix  $\Delta(U,V)=\left[\begin{smallmatrix}
U & V \\
V^\# & U^\#
\end{smallmatrix}\right]$. $I_k$ is the identity matrix of dimension $k$. Let $J_k={\rm diag}\{I_k,-I_k\}$ and  $\mathbb{J}_k=\bigl[
\begin{smallmatrix}
0_{k} & I_k \\
-I_k & 0_{k}%
\end{smallmatrix}
\big]$, define the $\flat$-adjoint and the $\sharp$-adjoint of $X\in\mathbb{C}^{2k\times2r}$ as  $X^\flat=J_rX^\dagger J_k$ and $X^\sharp=-\mathbb{J}_r X^\dagger\mathbb{J}_k$, respectively. $\delta_{jk}$ denotes the Kronecker delta function, $\delta(t-r)$ is the Dirac delta function, and $\otimes$ represents the tensor product. The commutator $[a,b]=ab-ba$ for operators $a,b$.  Finally, the reduced Planck constant $\hslash$ is set to be $1$.

\section{Linear quantum systems}\label{Preli}

\subsection{Modeling} \label{subsec:model}

The mathematical model of a linear quantum system composed of $n$ quantum harmonic oscillators interacting with $m$ input boson fields is briefly introduced in this subsection.  The $j$th quantum harmonic oscillator is described by the annihilation operator $\mbf{a}_j$ and its adjoint (the  creation operator) $\mbf{a}_j^\ast$; they satisfy the canonical commutation relations $[\mbf{a}_j,\mbf{a}_k^\ast]=\delta_{jk}$, $j,k=1,2,\ldots,n$. Stack annihilation operators into a column vector $
\mbf{a}=\left[\mbf{a}_1,\ldots,\mbf{a}_n\right]^\top
$. The $j$th input field is represented by an annihilation operator $\mbf{b}_{{\rm in},j}(t)$ and its adjoint $\mbf{b}_{{\rm in},j}^\ast(t)$, which enjoy  singular commutation relations:
\begin{equation*}
[\mbf{b}_{{\rm in},j}(t),\mbf{b}_{{\rm in},k}^\ast(r)]=\delta_{jk}\delta(t-r), ~ \forall j,k=1,2,\ldots,m, ~ t,r\in\mathbb{R}.
\end{equation*}
Denote the input vector $
\mbf{b}_{\rm in}(t)=\left[\mbf{b}_{{\rm in},1}(t),\ldots,\mbf{b}_{{\rm in},m}(t)\right]^\top
$.
It is convenient to characterize Markovian  quantum systems by the $(S,\mbf{L},\mbf{H})$ formalism  \cite{GJ09,GJ09b,CKS17,ZD22}. Here,  $S$ is a scattering operator satisfying $S^\dag S = S S^\dag = I_m$, the operators coupling the system to the input fields are represented by $\mbf{L}=\left[\begin{smallmatrix}
	C_- & C_+
\end{smallmatrix}\right]\breve{\mbf{a}}$ with $C_-, C_+\in\mathbb{C}^{m\times n}$, and the intrinsic system Hamiltonian  $\mbf{H}=\frac{1}{2}\breve{\mbf{a}}^\dagger\Omega\breve{\mbf{a}}$, where $\Omega=\Delta(\Omega_-,\Omega_+)$ is Hermitian with $\Omega_-, \Omega_+\in\mathbb{C}^{n\times n}$. In this paper, we assume that $S=I_m$ for simplicity.

In terms of the above system parameters, namely $\Omega_{\pm}$ and $ C_\pm$, the  Heisenberg equations of motion of a linear quantum system in the \emph{annihilation-creation operator representation} are
\begin{equation}\label{zerosys}\begin{aligned}
\dot{\breve{\mbf{a}}}(t)&=\mathcal{A}\breve{\mbf{a}}(t)+\mathcal{B}\breve{\mbf{b}}_{\rm in}(t), \\
\breve{\mbf{b}}_{\rm out}(t)&=\mathcal{C}\breve{\mbf{a}}(t)+\mathcal{D}\breve{\mbf{b}}_{\rm in}(t),
\end{aligned}\end{equation}
where the output vector $\mbf{b}_{\rm out}(t)=\left[\mbf{b}_{{\rm out},1}(t),\ldots,\mbf{b}_{{\rm out},m}(t)\right]^\top
$, and the complex-domain system matrices
\begin{equation}\label{eq:sys_matrices_ABCD}\begin{aligned}
& \mathcal{C}=\Delta(C_-,C_+), ~~ \mathcal{B}=-\mathcal{C}^\flat \mathcal{D}, \\
&\mathcal{A}=-\imath J_n\Omega-\frac{1}{2}\mathcal{C}^\flat \mathcal{C}, \ \mathcal{D}=I_{2m}.
\end{aligned}\end{equation}
The corresponding transfer matrix
\beq\label{eq:G_complex}
G(s) =  \mathcal{D} + \mathcal{C}(sI-\mathcal{A})^{-1}\mathcal{B}
\eeq
is assumed to be \emph{irreducible}, i.e., each entry of  $G(s)$ is irreducible with the same polynomial in $s$, that is, there are no hidden modes within each entry.

Taking the quantum expectation on both sides of Eq. \eqref{zerosys} with respect to the joint system-field initial state (\cite[Section 6.4.1]{WM10}, \cite[Section 2.6]{NY17}, \cite{NJP09}, \cite[Section 1]{BQD24}), yields a \emph{classical} linear system for mean dynamics
\beq \la{eq:real_sys_mean}
\begin{aligned}
\frac{d\l \breve{\mbf{a}}(t) \r}{dt} =& \mathcal{A} \l\breve{\mbf{a}}(t)\r + \mathcal{B} \l \breve{\mbf{b}}_{\rm in}(t)\r,
 \\
\l\breve{\mbf{b}}_{\rm out}(t)\r =& \mathcal{C} \l\breve{\mbf{a}}(t) \r+ \mathcal{D} \l \breve{\mbf{b}}_{\rm in}(t)\r.
\end{aligned}
\eeq
Thus, we can define controllability, observability, Hurwitz stability, detectability and  stabilizability for the linear \emph{quantum} system \eqref{zerosys} in terms of those for the linear \emph{classical} system \eqref{eq:real_sys_mean}.

\begin{definition}
\label{def:stab_ctrb_obsv}  (\cite[Def. 1]{GZ15} and \cite[Def. 3.1]{ZD22})
The linear quantum system \eqref{zerosys}  is said to be Hurwitz stable (resp. controllable, observable, detectable, stabilizable) if the corresponding linear classical system \eqref{eq:real_sys_mean} is Hurwitz stable (resp. controllable, observable, detectable, stabilizable).
\end{definition}

More discussion of open quantum systems can be found in, e.g.,  \cite{GZ00,JNP08,WM08,GJ09,WM10,CKS17,ZD22} and references therein.

\vspace{-2ex}
\subsection{The quantum Kalman canonical form}

\begin{figure}
  \centering
  \includegraphics[width=0.35\textwidth]{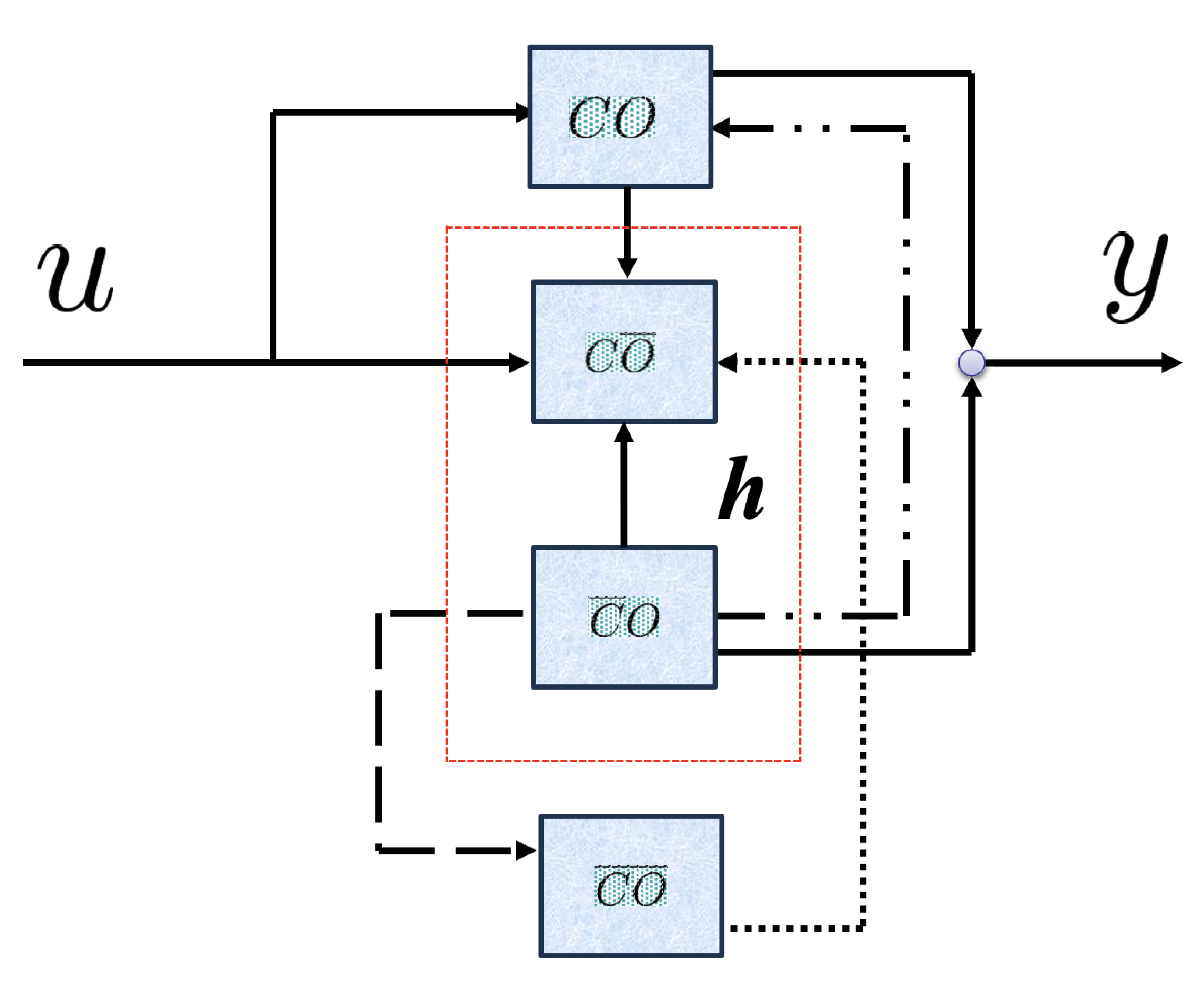}\\
  \caption{The Kalman canonical form of a linear quantum system; see \cite[Fig.~2]{ZGPG18}. The dimensions of the ``$co$'', ``$\bar{c}\bar{o}$'', and ``$h$'' (composed of the ``$c\bar{o}$'' subsystem and ``$\bar{c}o$'' subsystem) subsystems are $2n_1$, $2n_2$, $2n_3$, respectively, where $n_1,n_2,n_3\geq 0$ and $n_1+n_2+n_3=n$. ($n_k=0$ implies the absence of the corresponding subsystem.)
  } \label{KD}
\end{figure}

Based on controllability and observability defined in Definition \ref{def:stab_ctrb_obsv}, a special \emph{real-quadrature operator representation} of the linear quantum system \eqref{zerosys} was proposed in \cite[Thm. 4.4]{ZGPG18} via the Kalman decomposition, see Fig. \ref{KD} for the system diagram, which is
\begin{equation}\label{Kalmansys}\begin{aligned}
\dot{\mbf{x}}(t)&=\bar{A}\mbf{x}(t)+\bar{B}\mbf{u}(t), \\
\mbf{y}(t)&=\bar{C}\mbf{x}(t)+\bar{D}\mbf{u}(t),
\end{aligned}\end{equation}  with coordinates transformation \cite[Lem. 4.8 and Thm. 4.4]{ZGPG18}
\begin{equation}\label{eq:observables}\begin{aligned}
&\mbf{x}=\hat{T}^\dagger \breve{\mbf{a}}=\left[\begin{array}{c}
\mbf{q}_h \\
\mbf{p}_h \\ \hline
\mbf{x}_{co} \\ \hline
\mbf{x}_{\bar{c}\bar{o}}
\end{array}\right],
\\
&\mbf{u}=V_m\breve{\mbf{b}}_{\rm in}=\left[\begin{array}{c}
\mbf{q_{\rm in}} \\
\mbf{p_{\rm in}}
\end{array}\right], ~ \mbf{y}=V_m\breve{\mbf{b}}_{\rm out}=\left[\begin{array}{c}
\mbf{q_{\rm out}} \\
\mbf{p_{\rm out}}
\end{array}\right],
\end{aligned}\end{equation}
where the unitary matrix
\beqm
V_k=\frac{1}{\sqrt{2}}\left[\begin{array}{cc}
I_k & I_k \\
-\imath I_k & \imath I_k
\end{array}\right], \ \ k \in \mathbb{Z}^+,
\eeqm
and  $\hat{T}$ is unitary and also  preserves the canonical commutation relation $\imath \mathbb{\overline{J}}_{n} = [\mbf{x},\mbf{x}^\top] = \hat{T}^\dagger [\breve{\mbf{a}}, \breve{\mbf{a}}^\dagger]\hat{T}  =\hat{T}^\dagger  J_n\hat{T} $ with
\beq \label{eq:J_n_mbf}
 \mathbb{\overline{J}}_{n}
\triangleq
\left[
\begin{array}{ccc}
\mathbb{J}_{n_{3}} & 0 & 0 \\
0 & \mathbb{J}_{n_{1}} & 0 \\
0 & 0 & \mathbb{J}_{n_{2}}
\end{array}
\right].
\eeq

As given in \cite{ZGPG18,ZPL20}, the real system matrices in system \eqref{Kalmansys}  are
\begin{equation}\label{eq:july5_ABCD}\begin{aligned}
  \bar{A}=&\ \hat{T}^\dagger \mathcal{A} \hat{T}  = \mathbb{\overline{J}}_{n}\mathbb{H}+\frac{1}{2}\bar{B}\bar{C}, \\
  \bar{B}=&\ \hat{T}^\dagger \mathcal{B} V_m^\dagger = \mathbb{\overline{J}}_{n}\bar{C}^{\top}\mathbb{J}_{m}, \\
  \bar{C}=&\ V_m \mathcal{C}\hat{T}, \
  \bar{D} = V_m \mathcal{D} V_m^\dagger=I_{2m},
\end{aligned}\end{equation}
where the real symmetric matrix
\begin{equation}\label{J_n}
\mathbb{H} =\hat{T}^\dagger \Omega \hat{T}.
\end{equation}

 We denote the transfer matrix of system \eqref{Kalmansys} by $\mathbb{G}(s)$. Clearly,
\beq\label{eq:G_real}
    \mathbb{G}(s) = V_m G(s)V_m^\dag,
\eeq
which satisfies
\beq \la{eq:jun7_temp2}
\mathbb{G}(-s^\ast)^\sharp \mathbb{G}(s)=
\mathbb{G}(s)\mathbb{G}(-s^\ast)^\sharp= I_{2m},
\eeq
see, e.g.,  \cite{GJN10,ZJ13,GZ15} for more details.

As given in \cite[Eq. (67)]{ZGPG18}, the real system matrices $\bar{A},\bar{B},\bar{C}$ in Eq. \eqref{eq:july5_ABCD} are of block-wise structure, corresponding to the partition of system variables in Eq. \eqref{eq:observables}, as well as the subsystem blocks in the system diagram in Fig. \ref{KD}. More specifically,
\begin{equation}\label{real_Kalman_sys_A}
\begin{aligned}
\bar{A}
= &
\left[
\begin{array}{cc|c|c}
A_{h}^{11} & A_{h}^{12} & A_{12} & A_{13} \\
0 & A_{h}^{22} & 0 & 0  \\ \hline
0 & A_{21} & A_{co} & 0  \\ \hline
0 & A_{31} & 0 & A_{\bar{c}\bar{o}}
\end{array}
\right] ,  \ \ \bar{B}
=
\left[
\begin{array}{c}
B_{h} \\
0 \\ \hline
B_{co} \\ \hline
0
\end{array}
\right],
\\
\bar{C}
= &
\left[
\begin{array}{cc|c|c}
0 & C_{h} & C_{co} & 0
\end{array}
\right], \ \ \bar{D}  = I_{2m}.
\end{aligned}
\end{equation}

It can be readily verified that the system matrices in Eq. \eqref{eq:july5_ABCD} satisfy the so-called \emph{physical realizability conditions} for the linear quantum system \eqref{Kalmansys} \cite{ZGPG18,ZPL20}:
\begin{equation}\label{eq:phy_real}
\bar{A}\mathbb{\overline{J}}_{n}+\mathbb{\overline{J}}_{n}\bar{A}^{\top}+\bar{B}\mathbb{J}
_{m}\bar{B}^{\top}
=
0, \ \
\bar{B}
=
\mathbb{\overline{J}}_{n}\bar{C}^{\top}\mathbb{J}_{m},
\end{equation}
whose counterparts in the annihilation-creation operator representation can be derived by means of the system matrices in Eq. \eqref{eq:sys_matrices_ABCD}; see \cite[Eq. (4)]{ZGPG18} for details.

As mentioned in the Introduction, controllability and observability are equivalent for linear quantum systems \cite[Prop. 1]{GZ15}. Interestingly, however, detectability and stabilizability are not equivalent, as demonstrated  below.
\begin{example}\la{eq:Example_h}
In the $(S,\mbf{L},\mbf{H})$ formalism, let $S=1$,  $\mbf{H}=\frac{1}{2}\breve{\mbf{a}}^\dagger \Omega \breve{\mbf{a}}$, and $\mbf{L}=\left[\begin{smallmatrix}
C_- & C_+
\end{smallmatrix}\right]\breve{\mbf{a}}$, where $\Omega=\left[\begin{smallmatrix}
0 & -\imath \\
\imath & 0
\end{smallmatrix}\right]$ and $C_-=-\frac{\imath}{2}$, $C_+=\frac{\imath}{2}$. Then the resulting linear quantum system \eqref{Kalmansys} has
system matrices
\[
\bar{A}=\left[\bey{cc}
-1 & 0 \\
0 & 1
\eey
\right], \ \ \bar{B}=\bar{C}=\left[\bey{cc}
0 & 1 \\
0 & 0
\eey\right], \ \  \bar{D}=I_2.
\]
It can be readily shown that this linear quantum system has only the  ``$h$''   subsystem component. Moreover, it can be verified that it is detectable but not stabilizable.
\end{example}

More discussions on the quantum Kalman canonical form can be found in \cite{ZGPG18,GZPG17,ZPL20,ZLDP25}.


\subsection{Poles, invariant zeros and transmission zeros}

In this subsection, zeros and poles of linear quantum systems are defined.
Like  Hurwitz stability, controllability and observability defined in Definition \ref{def:stab_ctrb_obsv}, they are natural generalizations of their classical counterparts.

\bdfn (\cite[Def. 3.16]{ZDG96})
The \emph{invariant zeros of the linear quantum system realization} \eqref{Kalmansys} are the complex numbers $s_0$, which satisfy the inequality
\begin{equation}\label{eq:inv_zeros}
{\rm{rank}}(P(s_0))<{\rm normalrank}\; P \ \triangleq \max_{s\in\mathbb{C}} {\rm rank}(P(s)),
\end{equation}
where the Rosenbrock system matrix
\begin{equation}  \label{eq:P(s)}
P(s)\triangleq \left[\begin{array}{cc}\bar{A}-sI & \bar{B} \\
\bar{C} & \bar{D}\end{array}\right].
\end{equation}
\edfn

\begin{remark}\label{2n2m}
It is worth pointing out that
\beq \la{eq:normrank_P}
{\rm normalrank}\; P=2(n+m)
\eeq
 always holds for the open linear quantum system \eqref{Kalmansys} due to $\bar{D}=I_{2m}$. Clearly, the same is true   for system \eqref{zerosys}.
\end{remark}


\bdfn (\cite[Ch. 2.5]{Rosenbrock1970},\cite[Eqs. (6,7)]{SS89}) \label{dfn:decoupling zerso} For the linear quantum system realization  \eqref{Kalmansys},
if a complex number $s_0$  satisfies
\beq \la{eq:inv_zero}
{\rm rank}\left[
\bey{c}
\bar{A}-s_0I\\
\bar{C}
\eey
\right]<2n,
\eeq
then it is an unobservable mode (also called \emph{output-decoupling zero}). On the other hand, if
\beq \la{eq:inv_zero_2}
{\rm rank}[\bar{A}-s_0I \ \bar{B}]<2n,
\eeq
then $s_0$ is an  uncontrollable mode  (also called \emph{input-decoupling zero}).
\edfn
\bmrk
Due to Eq. \eqref{eq:normrank_P},  input-coupling zeros and  output-decoupling zeros of the linear quantum system realization \eqref{Kalmansys} are   invariant zeros.
\emrk

Let $r$ be the number of the unobservable modes. Then  the Rosenbrock system matrix $P(s)$ can be re-written in the observability decomposition form, \cite[Theorem 3.8]{ZDG96}
\begin{equation}\label{Feb3-2}
\tilde{P}(s)=\left[\begin{array}{ccc}
\bar{A}_o-sI_{2n-r} & 0 & \bar{B}_o  \\
\bar{A}_{21} & \bar{A}_{\bar{o}}-sI_{r} & \bar{B}_{\bar{o}} \\
\bar{C}_o & 0 & \bar{D}
\end{array}\right],
\end{equation}
where by the linear quantum system realization  \eqref{Kalmansys} with system matrices \eqref{real_Kalman_sys_A}, we know that
\begin{equation}\label{eq:ABCD}\begin{aligned}
&\bar{A}_{o}=\left[\begin{array}{cc}
A_{co} & A_{21} \\
0 & A_{h}^{22}
\end{array}\right], ~~
\bar{A}_{\bar{o}}=\left[
\begin{array}{cc}
A_h^{11} & A_{13} \\
0 & A_{\bar{c}\bar{o}}
\end{array}\right], \\
&\bar{A}_{21}=\left[\begin{array}{cc}
A_{12} & A_h^{12} \\
0 & A_{31}
\end{array}\right], ~~
\bar{C}_o=\left[\begin{array}{cc}
C_{co} & C_h
\end{array}\right], \\
&\bar{B}_o=\left[\begin{array}{c}
B_{co} \\
0
\end{array}\right], ~~
\bar{B}_{\bar{o}}=\left[\begin{array}{c}
B_h \\
0
\end{array}\right].
\end{aligned}\end{equation}
Clearly, the observable and unobservable modes are the eigenvalues of matrices $\bar{A}_o$ and $\bar{A}_{\bar{o}}$, respectively. According to Eq. \eqref{Feb3-2}, the invariant zeros of the linear quantum system realization \eqref{Kalmansys} consist of the eigenvalues of the matrix $\bar{A}_{\bar{o}}$ (namely the output-decoupling zeros) and the invariant zeros of
the observable subsystem realization or equivalently
\begin{equation}
P_o(s)=\left[\begin{array}{cc}
\bar{A}_o-sI_{2n-r} & \bar{B}_o \\
\bar{C}_o & \bar{D}
\end{array}\right].
\end{equation}

Similar to Eq. \eqref{Feb3-2}, let $k$ be the number of the uncontrollable modes. We have the following controllability decomposition form, \cite[Theorem 3.6]{ZDG96}
\begin{equation}
\label{eq:tilde_P_jun10}\begin{aligned}
\bar{P}(s)=\left[\begin{array}{ccc}
\bar{A}_c-sI_{2n-k} & \bar{A}_{12} & \bar{B}_c \\
0 & \bar{A}_{\bar{c}}-sI_k & 0 \\
\bar{C}_c & \bar{C}_{\bar{c}} & \bar{D}
\end{array}\right],
\end{aligned}\end{equation}
where
\begin{equation}\label{Aug22-1}\begin{aligned}
&\bar{A}_{c}=\left[\begin{array}{cc}
A_{co} & 0 \\
A_{12} & A_{h}^{11}
\end{array}\right], ~~
\bar{A}_{\bar{c}}=\left[
\begin{array}{cc}
A_h^{22} & 0 \\
A_{31} & A_{\bar{c}\bar{o}}
\end{array}\right], \\
&\bar{A}_{12}=\left[\begin{array}{cc}
A_{21} & 0 \\
A_h^{12} & A_{13}
\end{array}\right], ~~
\bar{B}_c=\left[\begin{array}{c}
B_{co} \\
B_h
\end{array}\right], \\
&\bar{C}_c=\left[\begin{array}{cc}
C_{co} & 0
\end{array}\right], ~~
\bar{C}_{\bar{c}}=\left[\begin{array}{cc}
C_h & 0
\end{array}\right].
\end{aligned}\end{equation}
Similarly, the controllable and uncontrollable modes are the eigenvalues of matrices $\bar{A}_c$ and $\bar{A}_{\bar{c}}$, respectively. According to Eq. \eqref{eq:tilde_P_jun10}, the invariant zeros of the linear quantum system realization \eqref{Kalmansys} consist of the eigenvalues of the matrix $\bar{A}_{\bar{c}}$ (namely the input-decoupling zeros) and the invariant zeros of
the controllable subsystem realization or equivalently
\begin{equation}
P_c(s)=\left[\begin{array}{cc}
\bar{A}_c-sI_{2n-k} & \bar{B}_c \\
\bar{C}_c & \bar{D}
\end{array}\right].
\end{equation}

Proposition \ref{odzero} in the next section establishes that $r=k$; in other words, the number of output-decoupling zeros equals that of the input-decoupling zeros.

In control theory, the eigenvalues of the $A$-matrix are called the \emph{poles} of the associated state-space realization.

The following result is recalled to prepare for the definition of transmission zeros of linear quantum systems.

\begin{lemma}(\cite[Thm. 2.3]{Maciejowski89})\label{McMillan}
Let $G(s)$ be a rational matrix function of normal rank $\ell$. Then $G(s)$ can be transformed by a series of elementary row and column operations $U(s)$, $V(s)$ into a pseudo-diagonal rational matrix $M(s)$ of the form
\begin{equation}\label{Eq. Smith-McMillan}\begin{aligned}
M(s)={\rm diag}\left\{\frac{\alpha_1(s)}{\beta_1(s)},\frac{\alpha_2(s)}{\beta_2(s)},\cdots,\frac{\alpha_\ell(s)}{\beta_\ell(s)},0,\cdots,0\right\},
\end{aligned}\end{equation}
in which $U(s)$ and $V(s)$ are unimodular polynomial matrix functions \cite[Eq. (2.33)]{Maciejowski89}, and the  monic polynomials $\{\alpha_i(s),\beta_i(s)\}$ are coprime for each $i=1,\ldots, \ell$ and satisfy the divisibility properties
\begin{equation}
\alpha_i(s)|\alpha_{i+1}(s), ~~ \beta_{i+1}(s)|\beta_i(s), ~~ i=1,\ldots,\ell-1.
\end{equation}
In the control literature, $M(s)$ is commonly referred to as the Smith-McMillan form of $G(s)$; \cite{HJ84,Maciejowski89,ZDG96}.
\end{lemma}

There are various definitions of transmission zeros in the literature; see for example \cite{kailath1980,DCC1982,CD1991,ZDG96}. In this paper, we adopt the definition of transmission zeros given  in \cite{ZDG96}.
\bdfn (\cite[Definition 3.14]{ZDG96})
The \emph{transmission zeros} of a transfer matrix $G(s)$ are the roots of any one of the numerator polynomials of the Smith-McMillan form given in Lemma \ref{McMillan}.  Moreover, $s_0$ is called a \emph{blocking zero} if  $\alpha_1(s_0)=0$. In this case, $G(s_0)=0$.
\edfn

\bcor
The blocking zeros of the transfer matrix $\mathbb{G}(s)$ of the linear quantum system realization  \eqref{Kalmansys}  cannot be purely imaginary.
\ecor
\emph{Proof.} Suppose $\mathbb{G}(s)$ has a purely imaginary blocking zero $s_0=\imath\omega_0$, then $\mathbb{G}(s_0)=0$ and $\mathbb{G}(-s_0^\ast)=0$. Thus $\mathbb{G}(-s_0^\ast)^\sharp \mathbb{G}(s_0)=0$, which contradicts Eq. \eqref{eq:jun7_temp2}.

The poles of a transfer matrix are defined as follows.

\bdfn (\cite[Definition 3.13]{ZDG96}) \label{def:pole}
A complex number $s_0\in\mathbb{C}$ is called a \emph{pole} of a transfer  matrix $G(s)$ if it is a root of any one of the denominator polynomials $\beta_j(s)$ in the Smith-McMillan \eqref{Eq. Smith-McMillan} form of $G(s)$.
\edfn


\begin{remark}
By definition, transmission zeros are defined in terms of transfer matrices, while invariant zeros are defined in terms of state-space realizations. However, by  \cite[Corollary 3.35]{ZDG96} a transmission zero must be an invariant zero. Moreover, by \cite[Theorem 3.34]{ZDG96} the transmission zeros and invariant zeros are identical for minimal system realizations.
\end{remark}



\bmrk \label{rem:full_rank}
Clearly, for the open linear quantum system \eqref{zerosys},  ${\rm det}[G(s)]\not\equiv0$ always holds as $\mathcal{D}=I_{2m}$. The same also holds for the linear quantum system realization  \eqref{Kalmansys}.
\emrk


\section{Zeros and poles of linear quantum systems}\label{ZPLQS}

In this section, we study the relation between zeros and poles of linear quantum systems.

\vspace{-2ex}
\subsection{Output-decoupling zeros and input-decoupling zeros}\label{subsec:decoupling zeros}

\bprop\label{odzero}
 $s_0$ is an output-decoupling zero of the linear quantum system realization  \eqref{Kalmansys} if and only if $-s_0$ is an input-decoupling zero.
\eprop

\emph{Proof.} Let $s_0$ be an output-decoupling zero of the linear quantum system realization  \eqref{Kalmansys}, i.e., $s_0$ satisfies Eq. \eqref{eq:inv_zero}. Then clearly $s_0^\ast$ is also an output-decoupling zero. Let the associated vector be $x$, that is, $\bar{A}x = s_0^\ast  x$, and $\bar{C}x=0$. Then by Eq. \eqref{eq:july5_ABCD}, we have $\mathbb{\overline{J}}_n\mathbb{H} x= s_0^\ast x$ and $x^\dagger\bar{C}^\top=0$. Define $z=\mathbb{\overline{J}}_n x$. We have $z^\dag \bar{B} = x^\dagger \bar{C}^\top \mathbb{J}_m=0$ and $-s_0 z^\dag = -s_0 x^\dag \mathbb{\overline{J}}_n^\top = -(x^\dagger \bar{A}^\top)\mathbb{\overline{J}}_n^\top=
x^\dagger \mathbb{H} = z^\dag \bar{A}$. Thus, $-s_0$ satisfies Eq. \eqref{eq:inv_zero_2} and is therefore an input-decoupling zero. The converse can be established in a similar way. \hfill $\blacksquare$

Noticing that $s_0^\ast$ is an eigenvalue of $\mathcal{A}^\dagger$  if $s_0$ is an eigenvalue of $\mathcal{A}$, the following is an immediate consequence of Proposition \ref{odzero}.
\bcor\label{cor3.1}
 $s_0$ is an output-decoupling zero of the linear quantum system realization \eqref{zerosys} if and only if $-s_0^\ast$ is an input-decoupling zero.
\ecor

\subsection{Transmission zeros and poles}\label{subsec:Transmission}

We begin with the following result.

\begin{proposition}\label{prop_G and inv G}
$s_0$ is a pole of the transfer matrix  $\mathbb{G}(s)$ define in \eqref{eq:G_real} if and only if $-s_0^\ast$ is a transmission zero  of $\mathbb{G}(s)$.
\end{proposition}

\emph{Proof.}
By Lemma \ref{McMillan} and  Remark \ref{rem:full_rank}, $\mathbb{G}(s)$ can be transformed to its Smith-McMillan form
\begin{equation}\label{june4-1}
\hspace{-1ex} U(s)\mathbb{G}(s)V(s)=M(s)=
{\rm diag}\left\{\frac{\alpha_1(s)}{\beta_1(s)},\frac{\alpha_2(s)}{\beta_2(s)},\cdots,\frac{\alpha_{2m}(s)}{\beta_{2m}(s)}\right\}.
\end{equation}
That $s_0$ being a pole of $\mathbb{G}(s)$ implies that there exists a polynomial $\beta_i(s)$ satisfying $\beta_i(s_0)=0$. Thus $s_0$ is a transmission zero of $\mathbb{G}(s)^{-1}$; see e.g., \cite[Lemma 3.38]{ZDG96}. On the other hand, from
Eq. \eqref {june4-1} we have
\begin{equation} \label{eq:jun11_1}
\begin{aligned}
V(-s^\ast &)^\sharp  \mathbb{G}(-s^\ast)^\sharp U(-s^\ast)^\sharp =M(-s^\ast)^\sharp \\
=\ {\rm diag}\Bigg\{&\frac{\alpha_{m+1}^\ast(-s^\ast)}{\beta_{m+1}^\ast(-s^\ast)},\cdots,\frac{\alpha_{2m}^\ast(-s^\ast)}{\beta_{2m}^\ast(-s^\ast)}, \\
&\frac{\alpha_{1}^\ast(-s^\ast)}{\beta_{1}^\ast(-s^\ast)}, \cdots,\frac{\alpha_{m}^\ast(-s^\ast)}{\beta_{m}^\ast(-s^\ast)}\Bigg\}.
\end{aligned}\end{equation}
By Eq. \eqref{eq:jun7_temp2}, $\mathbb{G}(s)^{-1} = \mathbb{G}(-s^\ast)^\sharp$. Consequently,  $s_0$ is also a transmission zero of $ \mathbb{G}(-s^\ast)^\sharp$ and therefore there must be a polynomial $\alpha_j^\ast(-s_0^\ast)=0$ and thus $\alpha_j(-s_0^\ast)=0$ in Eq. \eqref{eq:jun11_1}. As a result, $-s_0^\ast$ is a transmission zero. The converse  statement can be established in a similar way.  \hfill $\blacksquare$

Noticing that $s_0^\ast$ is a pole of $\mathbb{G}(s)$  if and only if $s_0$ is a pole of $\mathbb{G}(s)$,  Proposition \ref{prop_G and inv G} can be re-stated as follows.
\bthm\label{cor:s_-s}
$s_0$ is a pole of the transfer matrix  $\mathbb{G}(s)$ if and only if $-s_0$ is a transmission zero  of $\mathbb{G}(s)$.
\ethm

For the transfer matrix $G(s)$ defined in Eq. \eqref{eq:G_complex} in the complex domain, we have the following result. As the proof is similar to that of Proposition \ref{prop_G and inv G}, it is omitted.

\begin{corollary}\label{prop_G and inv G_real}
$s_0$ is a pole of the transfer matrix  $G(s)$ if and only if $-s_0^\ast$ is a transmission zero of $G(s)$.
\end{corollary}

\begin{remark} \label{rem:imaginary}
According to Proposition \ref{prop_G and inv G}, a purely imaginary pole is also a purely imaginary transmission zero of a linear quantum transfer matrix and vice versa.
\end{remark}

The following example demonstrates that the correspondence between transmission zeros and poles given in Theorem \ref{cor:s_-s} cannot be used to determine whether a system is quantum or not.
\begin{example}\label{exam3.2}
Consider a linear system with transfer matrix
\begin{equation}\label{July16-1}
\mathbb{G}(s)=\left[\begin{array}{cc}
\frac{s-1}{s+1} & 1 \\
0 & \frac{s+1}{s-1}
\end{array}\right],
\end{equation}
which satisfies Theorem \ref{cor:s_-s}. However, it can be verified that Eq. \eqref{eq:jun7_temp2} does not hold for this $\mathbb{G}(s)$.
Thus, the non-commutativity of a linear quantum system gives rise to the correspondence between system poles and zeros as characterized by Theorem \ref{cor:s_-s}; but on the other hand,  systems having such properties are not necessarily valid quantum-mechanical systems.
\end{example}


\subsection{Invariant zeros and poles}\label{subsec:invariance}

The relations between invariant zeros and poles for linear quantum systems are studied in this subsection.

\bprop\label{Feb3-3}
$s_0$ is an eigenvalue of $\mathcal{A}$ if and only if $-s_0^\ast$ is an invariant zero of the linear quantum system realization \eqref{zerosys}. Thus, the set of invariant zeros of the linear quantum system realization \eqref{zerosys} is  $\lambda(-\mathcal{A}_o^\dagger)\cup\lambda(-\mathcal{A}_{\bar{o}}^\dagger)$.
\eprop

\emph{Proof.} $s_0$ is an eigenvalue of $\mathcal{A}$ if and only if $s_0^\ast$ is an eigenvalue of $\mathcal{A}^\dagger$. Notice that
\begin{equation}\label{RSM2}
{\rm det}[s_0^\ast I-\mathcal{A}^\flat]
= {\rm det} \left[s_0^\ast I-\mathcal{A}^\dagger\right]=0.
\end{equation}
 Thus $s_0^\ast$ is also an eigenvalue of $\mathcal{A}^\flat$. Since $\mathcal{D}$ is unitary, it is easy to verify that the following identity holds
\begin{equation}\label{RSM1}\begin{aligned}
\left[\begin{array}{cc}\mathcal{A}-sI & \mathcal{B} \\ \mathcal{C} & \mathcal{D}\end{array}\right]
\left[\begin{array}{cc}I & 0 \\ -\mathcal{D}^{-1}\mathcal{C} & I\end{array}\right]=
\left[\begin{array}{cc}\mathcal{A}-sI+\mathcal{C}^\flat\mathcal{C} & \mathcal{B} \\ 0 & \mathcal{D}\end{array}\right],
\end{aligned}\end{equation}
where the second equation in Eq. \eqref{eq:sys_matrices_ABCD}
has been used in the derivation.
By Eqs. \eqref{eq:july5_ABCD} and \eqref{RSM1},
\begin{equation}\label{Feb3-1}\begin{aligned}
&{\rm det}[P(s)]={\rm det}\left[\mathcal{A}-sI+\mathcal{C}^\flat\mathcal{C}\right] \\
=\ &{\rm det}\left[-sI-\imath J_n\Omega+\frac{1}{2}\mathcal{C}^\flat\mathcal{C}\right]={\rm det}\left[sI+\mathcal{A}^\flat\right].
\end{aligned}\end{equation}
Let $s=-s_0^\ast$ in Eq. \eqref{Feb3-1}. By Eq. \eqref{RSM2} we have ${\rm det}[P(-s_0^\ast)]=0$, which means that $-s_0^\ast$ must be an invariant zero of the linear quantum system realization \eqref{zerosys}. Conversely, if $s_0$ is an invariant zero of the linear quantum system realization \eqref{zerosys}, then by  Eq. \eqref{Feb3-1} ${\rm det}[P(s_0)]=0$ implies $s_0$ is an eigenvalue of $-\mathcal{A}^\flat$. Thus, $-s_0^\ast$ is an eigenvalue of $\mathcal{A}$.  \hfill $\blacksquare$

For the linear quantum system realization  \eqref{Kalmansys} in the real domain, we have the following result, whose proof is similar to  that of Proposition \ref{Feb3-3} and thus is omitted.

\bthm\label{cor:invariant_0}
$s_0$ is an eigenvalue of the matrix $\bar{A}$ if and only if $-s_0$ is an invariant zero of the linear quantum system realization \eqref{Kalmansys}. Thus, the set of invariant zeros of the linear quantum system realization \eqref{Kalmansys} is  $\lambda(-\bar{A}_o)\cup \lambda(-\bar{A}_{\bar{o}})$.
\ethm

\begin{example}\label{ex:3.2}
Consider a linear system with system matrices $A=\left[\begin{smallmatrix}
1 & 0 \\
0 & -1
\end{smallmatrix}\right]$, $B=\left[\begin{smallmatrix}
0 & 0 \\
1 & 1
\end{smallmatrix}\right]$, $C=\left[\begin{smallmatrix}
1 & 0 \\
1 & 0
\end{smallmatrix}\right]$, and $D=I_2$. Both the eigenvalues of $A$ and the invariant zeros of this system realization are $\pm1$. However, it can be verified that Eq.  \eqref{eq:jun7_temp2} does not hold  for this system. Thus, it is not a valid quantum-mechanical system. As a result, the correspondence between eigenvalues of the $A$-matrix  and invariant zeros of a system realization given in Theorem \ref{cor:invariant_0} cannot be used to determine whether a system is quantum or not.
\end{example}

\begin{remark} \label{rem:1-to-1}
Obviously, the set of poles of a transfer matrix is a subset of the eigenvalues of  the $A$-matrix of a state-space realization, while the set of transmission zeros of a transfer matrix is a subset of invariant zeros of a state-space realization. Loosely speaking, Theorem  \ref{cor:invariant_0} is a generalization of Theorem  \ref{cor:s_-s}. Thus, in the linear quantum regime, there exist a duality between  poles and transmission zeros of a transfer matrix and another duality  between eigenvalues of the $A$-matrix  and invariant zeros of a system realization.
\end{remark}

\begin{assumption}\label{Feb-h} For  the Kalman canonical form  \eqref{Kalmansys}, we assume that the poles of the ``$h$"  subsystem and the eigenvalues of the  matrix $A_{\bar{c}\bar{o}}$   are purely imaginary.
\end{assumption}

\begin{remark}
Firstly, Assumption \ref{Feb-h} holds for linear \emph{passive} quantum systems, see \cite{ZGPG18, ZPL20}. Secondly, if the linear quantum system \eqref{Kalmansys} is both controllable and observable, namely it is a minimal realization, then it has
neither the ``$h$" subsystem nor the ``$\bar{c}\bar{o}$" subsystem (as shown in Fig. \ref{KD});  therefore
Assumption 3.1 \emph{naturally} holds for this class of linear quantum systems. Thirdly, as Hurwitz stability implies both controllability and observability \cite[Thm. 3.1]{ZPL20},  Assumption \ref{Feb-h} holds  for stable linear quantum systems. Fourthly, many \emph{physical} systems satisfy Assumption \ref{Feb-h}; see for example \cite{DFK+12,NY14,LOW+21}. Finally, nevertheless there are indeed linear quantum systems whose ``$h$'' subsystems do not satisfy Assumption \ref{Feb-h}; see Example \ref{eq:Example_h} for an illustration.
\end{remark}

Under Assumption \ref{Feb-h}, $\lambda(A_{h}^{11})=\lambda(-A_{h}^{11})$ and $  \lambda(A_{h}^{22})= \lambda(-A_{h}^{22})$. Moreover, by \cite[Lem. 3.1]{ZPL20},   $A^{11}_h = - A^{22^\top}_h$. Therefore, under Assumption \ref{Feb-h} Theorem \ref{cor:invariant_0} can be refined as follows.
\bthm\label{cor:union_eig}
Under Assumption \ref{Feb-h}, the set of invariant zeros of the linear quantum system realization \eqref{Kalmansys} is  $\lambda(-A_{co})\cup \lambda(A_{\bar{c}\bar{o}})\cup \lambda(A_{h}^{11}) $.
\ethm

In what follows, we take a close look at  Assumption \ref{Feb-h}. According to Eqs. \eqref{Kalmansys}, \eqref{eq:observables}, and \eqref{real_Kalman_sys_A}, the evolutions of the system variables ``$\mbf{x}_{\bar{c}\bar{o}}$'' and ``$\mbf{p}_h$'' are not affected by the inputs either directly or indirectly,  thus one may wonder whether they are isolated systems. If so, then they evolve unitarily and consequently, all the eigenvalues of the matrices $A_{\bar{c}\bar{o}}$ and $A^{22}_h$  must be purely imaginary. Then all the eigenvalues of the matrix $A^{11}_h$ are also purely imaginary as $A^{11}_h = - A^{22^\top}_h$ \cite[Lem. 3.1]{ZPL20}. That is, Assumption \ref{Feb-h} naturally holds. Unfortunately, the above is not true, and  in some instances the uncontrollable and unobservable  (``$\bar{c}\bar{o}$'')  subsystem and the ``$\mbf{p}_h$''  variable are not isolated systems.  We take the ``$\bar{c}\bar{o}$'' subsystem as an example. For the ``$\bar{c}\bar{o}$'' subsystem, from \cite[Lem. 3.3]{ZPL20}  the matrix  $A_{\bar{c}\bar{o}}$ indeed contains no contribution from the input fields; instead, it is completely determined by the intrinsic system Hamiltonian $\mbf{H}$.  In the annihilation-creation operator representation, as discussed in Subsection \ref{subsec:model},    $\mbf{H}=\frac{1}{2}\breve{\mbf{a}}^\dagger\Omega\breve{\mbf{a}}$, where $\Omega=\Delta(\Omega_-,\Omega_+)$ is Hermitian with $\Omega_-, \Omega_+\in\mathbb{C}^{n\times n}$.  The existence of the  $\Omega_+$ term means there is energy input (often called pump in quantum optics) to the system. In other words, in the mathematical modeling,  the contribution of the pump is often modeled as part of the intrinsic system Hamiltonian $\mbf{H}$ which is represented by the term $\Omega_+$, instead an explicit quantum input channel. This is the so-called semi-classical approximation. Thus in reality, the presence of $\Omega_+$  in the intrinsic system Hamiltonian $\mbf{H}$ indicates that the system is not isolated, and thus it does not evolve unitarily and accordingly the eigenvalues of the matrix $A_{\bar{c}\bar{o}}$  may not be purely imaginary. An example is given below.

\begin{example}\label{ex:bar_co}
Consider a linear quantum system  of the form
\begin{equation}\begin{aligned}
\dot{\mbf{x}}_{co}&=\left[\begin{array}{cc}
-\frac{\kappa}{2} & \omega \\
-\omega & -\frac{\kappa}{2}
\end{array}\right]\mbf{x}_{co}-\sqrt{\kappa}\ \mbf{u}, \\
\dot{\mbf{x}}_{\bar{c}\bar{o}}&=\left[\begin{array}{cc}
-2 & 1 \\
 -1 & 2
\end{array}\right]\mbf{x}_{\bar{c}\bar{o}}, \\
\mbf{y}&=\sqrt{\kappa}\ \mbf{x}_{co}+\mbf{u}.
\end{aligned}\end{equation}
The $(S,\mbf{L},\mbf{H})$ parameters for this system are $C_-=\left[\begin{smallmatrix}
\sqrt{\kappa} & 0
\end{smallmatrix}\right]$, $C_+=\left[\begin{smallmatrix}
0 & 0
\end{smallmatrix}\right]$, $\Omega_-=\left[\begin{smallmatrix}
\omega & 0 \\
0 & 1
\end{smallmatrix}\right]$, and $\Omega_+=\left[\begin{smallmatrix}
0 & 0 \\
0 & -2\imath
\end{smallmatrix}\right]$.
Although  decoupled from the ``$co$" subsystem, due to the existence of $\Omega_+$, the ``$\bar{c}\bar{o}$" subsystem  has eigenvalues $\pm\sqrt{3}$, which are not purely imaginary.
\end{example}

The following result gives a sufficient condition under which the eigenvalues of the matrix $A_{\bar{c}\bar{o}}$ are purely imaginary.

\bprop\label{prop3.4}
Let $\lambda$ be an eigenvalue of $A_{\bar{c}\bar{o}}$ and $x$ be a corresponding eigenvector. If $ x^\dag \mathbb{J}x\neq 0$, then  $\lambda$ is purely imaginary.
\eprop

\emph{Proof.} According to Eq. \eqref{Kalmansys},
\beq
\boldsymbol{\dot{x}}_{\bar{c}\bar{o}} =
A_{\bar{c}\bar{o}}\boldsymbol{x}_{\bar{c}\bar{o}}+A_{31}\boldsymbol{p}_{h}.
\la{sep6_1d}
\eeq
Integrating both sides of  Eq. \eqref{sep6_1d}
yields
\beqn
\boldsymbol{x}_{\bar{c}\bar{o}}(t)
= e^{A_{\bar{c}\bar{o}}t} \boldsymbol{x}_{\bar{c}\bar{o}}(0)
+ \int_0^t e^{A_{\bar{c}\bar{o}}(t-\tau)} A_{31}\boldsymbol{p}_{h}(\tau)d\tau.
\eeqn
Thus,
\begin{align}
&\imath \mathbb{J}_{n_2}= [\boldsymbol{x}_{\bar{c}\bar{o}}(t) ,  \ \ \boldsymbol{x}_{\bar{c}\bar{o}}(t)^\top]
\nonumber\\
=&\;
e^{A_{\bar{c}\bar{o}}t} [\boldsymbol{x}_{\bar{c}\bar{o}}(0) ,  \ \ \boldsymbol{x}_{\bar{c}\bar{o}}(0)^\top]e^{A_{\bar{c}\bar{o}}^\top t}
\nonumber
\\
&+ e^{A_{\bar{c}\bar{o}}t} \int_0^t [\boldsymbol{x}_{\bar{c}\bar{o}}(0),  \ \ \boldsymbol{p}_{h}(\tau)^\top]
A_{31}^\top e^{A_{\bar{c}\bar{o}}^\top(t-\tau)} d\tau
\nonumber
\\
 &+\int_0^t e^{A_{\bar{c}\bar{o}} (t-\tau)}A_{31}
  [\boldsymbol{p}_{h}(\tau),  \ \ \boldsymbol{x}_{\bar{c}\bar{o}}(0)^\top] e^{A_{\bar{c}\bar{o}}^\top t} d\tau
\nonumber
\\
 &+\int_0^t \int_0^t e^{A_{\bar{c}\bar{o}} (t-\tau)}A_{31}
  [\boldsymbol{p}_{h}(\tau), \boldsymbol{p}_{h}(r)^\top] A_{31}^\top   e^{A_{\bar{c}\bar{o}}^\top (t-r)} d\tau dr
  \nonumber
\\
=&\; \imath e^{A_{\bar{c}\bar{o}}t} \mathbb{J}_{n_2} e^{A_{\bar{c}\bar{o}}^\top t},
\label{eqLjuly10_temp1}
\end{align}
where the fact that  $[\boldsymbol{p}_{h}(\tau),\boldsymbol{p}_{h}(r)^{\top}]=0$ for all $0
\leq \tau, r\leq t$ is used. As a result,  $I = (e^{A_{\bar{c}\bar{o}}t} \mathbb{J}_{n_2}) ( \mathbb{J}_{n_2} e^{A_{\bar{c}\bar{o}}t})^\top$.
Let $\lambda$ be an eigenvalue of the matrix $A_{\bar{c}\bar{o}}$, clearly it is also an eigenvalue of the matrix $A_{\bar{c}\bar{o}}^\top$. Denote $x$ an associated eigenvector, then pre-multiplying and post-multiplying Eq. \eqref{eqLjuly10_temp1}  by $x^\dag$ and $x$ respectively yield  $x^\dag \mathbb{J}_{n_2} x = e^{2{\rm Re}(\lambda)t} x^\dag\mathbb{J}_{n_2} x $. Consequently, if $ x^\dag \mathbb{J}_{n_2} x\neq 0$, then $\lambda$ is purely imaginary. \hfill $\blacksquare$

We end this section with a remark on the eigenvalues of the  ``$h$'' subsystem.
\bmrk \label{rem:h_imaginary}
By  \cite[Lem. 3.1]{ZPL20},  the eigenvalues of the ``$h$'' subsystem are those of the Hamiltonian matrix
\beq
\left[
\begin{array}{cc}
A_{h}^{11} & 0 \\
0 & A_{h}^{22}
\end{array}
\right] =
\left[\begin{array}{cc}
-A_{h}^{22^\top} & 0 \\
0 & A_{h}^{22}
\end{array}
\right].
\eeq
Clearly, if the matrix $A_{h}^{22}$ is skew-symmetric, then all the eigenvalues of the ``$h$'' subsystem  are purely imaginary.
\emrk

\section{Invariant zeros and strong left invertibility}\label{IZLI}

In classical (namely, non-quantum mechanical) linear control literature, left invertibility is a powerful tool for feedforward control and learning control \cite{Moylan77,Zames81,Hautus83,HP98,Sogo10,CP18,DD2023, RO23,KT25}. Recently, it is proved in \cite{DD2023} that if a classical linear system is asymptotically strongly left invertible (see Definition \ref{def:invertibility} below), then there exists a \emph{stable inversion} such that the input to the original system  can be asymptotically reconstructed from the output. For an open linear quantum system $G$,  by Eq. \eqref{eq:jun7_temp2} the inverse $G^{-1}$ always exists; but there is no guarantee that $G^{-1}$ is stable.  By means of the pole-zero duality derived in the preceding section, in this section we study left invertibility of linear quantum systems, and based on which two types of stable input observers are constructed. It is worth noting that feedforward control is quite useful in the implementation of measurement-based  optical quantum computation \cite{LGB+21,SKH+23}.

\subsection{Left invertibility  of linear quantum  systems}\la{eq:subsec:left invertibility}

We first recall the definitions of  left invertibility for linear systems.

\begin{definition}\label{def:left_inv}(\cite[Def. 1]{DD2023})
A classical finite-dimensional linear time-invariant (FDLTI) system
\begin{equation}\label{eq:ss_july21}
\begin{aligned}
\dot{x}(t)&=Ax(t)+Bu(t), \\
y(t)&=Cx(t)+Du(t)
\end{aligned}\end{equation}
is said to be left invertible if for $x(0) = 0$,
\[
y(t) = 0 ~~ \text{for} ~~ t > 0 \Longrightarrow  u(t) = 0 ~~ \text{for} ~~ t > 0.
\]
\end{definition}

\begin{definition} \label{def:invertibility}(\cite[Def. 5]{DD2023})
The classical FDLTI system  \eqref{eq:ss_july21}
is said to be:
\begin{itemize}

\item strongly (s.-) left invertible if for any initial condition $x(0)$ and input $u(t)$,
\begin{equation*}
y(t)=0 ~~ \text{for} ~~ t>0 ~~~~ \Longrightarrow ~~~~ u(t)=0 ~~ \text{for} ~~ t>0;
\end{equation*}

\item asymptotically strongly (a.s.-) left invertible if for any initial condition $x(0)$ and input  $u(t)$,
\begin{equation*}
y(t)=0 ~~ \text{for} ~~ t>0 ~~~~ \Longrightarrow ~~~~ u(t)\longrightarrow0 ~~ \text{as} ~~ t\longrightarrow \infty;
\end{equation*}

\item asymptotically $\text{strong}^{\star}$ ($\text{a.s.}^{\star}$-) left invertible if for any initial condition $x(0)$ and input $u(t)$,
\begin{equation*}
y(t)\longrightarrow0 ~~ \text{as} ~~ t\longrightarrow\infty ~~~~ \Longrightarrow ~~~~ u(t)\longrightarrow0 ~~ \text{as} ~~ t\longrightarrow \infty.
\end{equation*}

\end{itemize}
\end{definition}

\bmrk \label{rem:inversion_detectability}
Clearly, strong left invertibility implies left invertibility.  Moreover, asymptotic strong (a.s.-) left invertibility  is equivalent to detectability defined in \cite[Def. 2]{HP98}. And correspondingly, Theorem 14 in \cite{DD2023} used in the proof of Corollary \ref{Thm6.1} below is equivalent to \cite[Thm. 2]{HP98}.
\emrk

\bmrk
Similar to Definition \ref{def:stab_ctrb_obsv}, we say that
the linear quantum system \eqref{zerosys} (resp.  the linear quantum system \eqref{Kalmansys}) is left invertible (resp. strongly (s.-) left invertible, asymptotically strongly (a.s.-) left invertible, asymptotically $\text{strong}^{\star}$ ($\text{a.s.}^{\star}$-) left invertible) if the linear classical system \eqref{eq:real_sys_mean} is left invertible (resp. strongly (s.-) left invertible, asymptotically strongly (a.s.-) left invertible, asymptotically $\text{strong}^{\star}$ ($\text{a.s.}^{\star}$-) left invertible).
\emrk

The following necessary and sufficient condition for a.s.-left invertibility of linear quantum systems is an immediate consequence of Theorem \ref{cor:invariant_0}.

\begin{corollary}\label{corollary4.1}
The linear quantum system \eqref{Kalmansys} is a.s.-left invertible if and only if $\lambda(-\bar{A}_c)/\lambda(\bar{A}_{\bar{o}})$ are in the left-half plane, or equivalently  $\{-\lambda: \lambda\in\lambda(\bar{A}_c), ~ {\rm Re}(\lambda)\leq0\}\subseteq\lambda(\bar{A}_{\bar{o}})$.
\end{corollary}
\emph{Proof.} By Theorem \ref{cor:invariant_0} and Eq. \eqref{eq:ABCD}, the set of invariant zeros of the linear quantum system realization \eqref{Kalmansys} is
\begin{equation}\begin{aligned}
		&\lambda(-\bar{A}_o)\cup \lambda(-\bar{A}_{\bar{o}}) \\
		=&\lambda(-A_{co})\cup\lambda(-A_h^{22})\cup\lambda(-A_h^{11})\cup\lambda(-A_{\bar{c}\bar{o}}).
\end{aligned}\end{equation}
By \cite[Lem. 3.1]{ZPL20}, $\lambda(-A_h^{22})=\lambda(A_h^{11})$, which implies that the unobservable modes $\lambda(\bar{A}_{\bar{o}})=\lambda(-A_h^{22})\cup\lambda(A_{\bar{c}\bar{o}})$. Moreover,  by \cite[Remark 3.1]{ZPL20}, $A_{\bar{c}\bar{o}}$ is a Hamiltonian matrix, therefore $\lambda(-A_{\bar{c}\bar{o}})=\lambda(A_{\bar{c}\bar{o}})$. Consequently, the subset of invariant zeros that are not output-decoupling zeros can be described as $(\lambda(-A_{co})\cup\lambda(-A_h^{11}))/\lambda(\bar{A}_{\bar{o}})=\lambda(-\bar{A}_c)/\lambda(\bar{A}_{\bar{o}})$. By \cite[Thm. 14]{DD2023} and Remark \ref{2n2m}, the linear quantum system \eqref{Kalmansys} is a.s.-left invertible if and only if the subset of invariant zeros of $P(s)$ that are not output-decoupling zeros is in the left-half plane, which means that $\lambda(-\bar{A}_c)/\lambda(\bar{A}_{\bar{o}})$ are in the left-half plane. In what follows, we proof that this condition is equivalent to $\{-\lambda: \lambda\in\lambda(\bar{A}_c), ~ {\rm Re}(\lambda)\leq0\}\subseteq\lambda(\bar{A}_{\bar{o}})$. As
$\{-\lambda: \lambda\in\lambda(\bar{A}_c), ~ {\rm Re}(\lambda)\leq0\}=\{\lambda: \lambda\in\lambda(-\bar{A}_c), ~ {\rm Re}(\lambda)\geq0\}$, we have $\{-\lambda: \lambda\in\lambda(\bar{A}_c), ~ {\rm Re}(\lambda)\leq0\}\subseteq\lambda(\bar{A}_{\bar{o}})$ is equivalent to
\begin{equation}\label{eq:lam}
\{\lambda: \lambda\in\lambda(-\bar{A}_c), ~ {\rm Re}(\lambda)\geq0\}\subseteq\lambda(\bar{A}_{\bar{o}}).
\end{equation}
If $\lambda\in\lambda(-\bar{A}_c)$, and $\lambda\notin\lambda(\bar{A}_{\bar{o}})$, i.e., $\lambda\in\lambda(-\bar{A}_c)/\lambda(\bar{A}_{\bar{o}})$, then Eq. \eqref{eq:lam} implies that ${\rm Re}(\lambda)<0$. Thus, $\lambda(-\bar{A}_c)/\lambda(\bar{A}_{\bar{o}})$ are in the left-half plane. Next, we show that $\lambda(-\bar{A}_c)/\lambda(\bar{A}_{\bar{o}})$ are in the left-half plane implies that Eq. \eqref{eq:lam} holds. By contradiction, if there exists $\lambda_0$ such that $\lambda_0\in\lambda(-\bar{A}_c)$, ${\rm Re}(\lambda_0)\geq0$, and $\lambda_0\notin\lambda(\bar{A}_{\bar{o}})$, then $\lambda_0\in\lambda(-\bar{A}_c)/\lambda(\bar{A}_{\bar{o}})$ but ${\rm Re}(\lambda_0)\geq0$, which conflicts with $\lambda(-\bar{A}_c)/\lambda(\bar{A}_{\bar{o}})$ being in the left-half plane. Consequently, the equivalence of the two conditions are established.  \hfill $\blacksquare$

Combining Corollary \ref{corollary4.1} with Assumption \ref{Feb-h} we have the following result.
\bthm\label{Thm6.1}
Under Assumption \ref{Feb-h}, the linear quantum system \eqref{Kalmansys}  is a.s.-left invertible if and only if the eigenvalue set $\lambda(A_{co})/ \lambda(\bar{A}_{\bar{o}})$ is in the right-half plane.
\ethm
\emph{Proof.}
By Corollary \ref{corollary4.1} and Eq. \eqref{Aug22-1}, the linear quantum system \eqref{Kalmansys} is a.s.-left invertible if and only if $(\lambda(-A_{co})\cup\lambda(-A_h^{11}))/\lambda(\bar{A}_{\bar{o}})$ are in the left-half plane, or equivalently, $(\lambda(A_{co})\cup\lambda(A_h^{11}))/\lambda(-\bar{A}_{\bar{o}})$ are in the right-half plane. Under Assumption \ref{Feb-h}, we have $\lambda(A_h^{11})=\lambda(A_h^{22})\subseteq\lambda(\bar{A}_{\bar{o}})=\lambda(-\bar{A}_{\bar{o}})$, and thus  the linear quantum system \eqref{Kalmansys} is a.s.-left invertible if and only if $\lambda(A_{co})/\lambda(\bar{A}_{\bar{o}})$ are in the right-half plane. \hfill $\blacksquare$

The following sufficient condition is an immediate consequence of Theorem \ref{Thm6.1}, whose proof is thus omitted.
\begin{corollary} \label{cor:A_co}
Under Assumption \ref{Feb-h}, the linear quantum system \eqref{Kalmansys} is a.s.-left invertible if the set of  the controllable and observable modes  $\lambda(A_{co})$ is in the right-half plane.
\end{corollary}

The following example demonstrates  the necessity of Assumption \ref{Feb-h}  in Theorem \ref{Thm6.1} as well as Corollary \ref{cor:A_co}.

\begin{example}\label{example_4.1}
Let the linear quantum system \eqref{Kalmansys} be
\begin{equation}\label{May17-1}\begin{aligned}
&\left[\begin{array}{c}
\dot{\mbf{q}}_h \\
\dot{\mbf{p}}_h \\ \hline
\dot{\mbf{x}}_{co}
\end{array}\right]=\bar{A}\left[\begin{array}{c}
\mbf{q}_h \\
\mbf{p}_h \\ \hline
\mbf{x}_{co}
\end{array}\right]+\bar{B}\left[\begin{array}{c}
\mbf{u}_1 \\
\mbf{u}_2
\end{array}\right], \\
&\left[\begin{array}{c}
\mbf{y}_1 \\
\mbf{y}_2
\end{array}\right]=\bar{C}\left[\begin{array}{c}
\mbf{q}_h \\
\mbf{p}_h \\ \hline
\mbf{x}_{co}
\end{array}\right]+\bar{D}\left[\begin{array}{c}
\mbf{u}_1 \\
\mbf{u}_2
\end{array}\right],
\end{aligned}\end{equation}
where $\mbf{x}_{co}=\left[\begin{array}{cc}
\mbf{x}_1 & \mbf{x_2}
\end{array}\right]^\top$, and
\begin{equation}\begin{aligned}
&\bar{A}=\left[\bey{cc|cc}
-1 & 0 & 1 & 0 \\
0 & 1 & 0 & 0 \\ \hline
0 & 0 & \frac{1}{2} & 0 \\
0 & 0 & 0 & \frac{1}{2}
\eey
\right], \ \ \bar{B}=\left[\bey{cc}
0 & 1 \\
0 & 0 \\ \hline
0 & 1 \\
1 & 0
\eey\right], \\
&\bar{C}=\left[\bey{cc|cc}
0 & 1 & 0 & 1 \\
0 & 0 & 1 & 0
\eey\right], \ \ \bar{D}=I_2.
\end{aligned}\end{equation}
Clearly, $\lambda(A_{co})=\{\frac{1}{2}\}$ and  $\lambda(\bar{A}_{\bar{o}})=\{-1\}$. Thus the set $\lambda(A_{co})/ \lambda(\bar{A}_{\bar{o}})=\lambda(A_{co})$ is in the right-half plane. Hence, the \emph{sufficiency}  part  of  both Theorem \ref{Thm6.1} and Corollary \ref{cor:A_co} holds, but Assumption \ref{Feb-h} does not.  On the other hand, $\mbf{y}_2(t)=0$ for $t>0$ yields that $\mbf{x}_1(t)=e^{-\frac{1}{2}t}\mbf{x}_1(0)$ and $\mbf{u}_2(t)=-\mbf{x}_1(t)=-e^{-\frac{1}{2}t}\mbf{x}_1(0)$ which converges to $0$ as $t\to0$. However, $\mbf{y}_1(t)=0$ for $t>0$ implies that $\mbf{x}_2(t)=e^{-\frac{1}{2}t}\mbf{x}_2(0)+\frac{2}{3}e^{-\frac{1}{2}t}\mbf{p}_h(0)-\frac{2}{3}e^t\mbf{p}_h(0)$ and $\mbf{u}_1(t)=-e^{-\frac{1}{2}t}\mbf{x}_2(0)-\frac{2}{3}e^{-\frac{1}{2}t}\mbf{p}_h(0)-\frac{1}{3}e^t\mbf{p}_h(0)$, which does not converge to $0$ for non-zero initial condition $\mbf{p}_h(0)$. Thus, the linear quantum system \eqref{May17-1} is not a.s.-left invertible.
\end{example}

\begin{remark}
Compared with \cite[Thm. 14]{DD2023}, Theorem \ref{Thm6.1} is derived with the aid of the  pole-zero duality of linear quantum systems. Such duality is not applicable to linear classical systems. Consider a \emph{classical} system with system matrices $A=\left[\begin{smallmatrix}
1 & 0 \\
0 & 0
\end{smallmatrix}\right]$, $B=\left[\begin{smallmatrix}
-1 \\
0
\end{smallmatrix}\right]$, $C=\left[\begin{smallmatrix}
1 & 0
\end{smallmatrix}\right]$, and $D=1$. Clearly,  both the \emph{sufficiency} part of Theorem \ref{Thm6.1} and  Assumption \ref{Feb-h} hold.  Let the state be $x(t)= [x_1(t) \ x_2(t)]^\top$.  It can be readily shown that $y=0$ implies that $u(t)=-e^{2t}x_1(0)$, which does not converge to zero asymptotically for any nonzero initials state $x_1(0)$. Thus, this classical linear system is not a.s.-left invertible. Actually, the subset of the invariant zeros of $P(s)$ that are not output-decoupling zeros is $\{2\}$, which does not satisfy the condition in \cite[Thm. 14]{DD2023}, and hence the system is not a.s.-left invertible.
\end{remark}

For minimal realizations we have the following result, which is an immediate consequence of Corollary \ref{corollary4.1}.

\begin{corollary}\label{cor:minimal}
Let the linear quantum system \eqref{Kalmansys} be a minimal realization. Then it is a.s.-left invertible if and only if  $\lambda(\bar{A}) =\lambda(A_{co}) $  is in the right-half plane.
\end{corollary}

\bmrk
Theorem \ref{Thm6.1} annd Corollaries \ref{cor:A_co} and  \ref{cor:minimal}  show that the ``$co$'' subsystem must be unstable in order that a linear quantum  system is a.s.- left invertible. This reveals the structure of linear \emph{quantum} systems, in particular their pole-zero duality.
\emrk

We conclude this subsection with a final result.

\bcor\label{corollary4.4}
The a.s.-left invertibility and $\text{a.s.}^{\star}$-left invertibility are equivalent  for linear quantum systems.
\ecor
\emph{Proof.} It follows from \cite[Thm. 15]{DD2023} and Remark \ref{2n2m} directly.

\vspace{-1ex}
\subsection{Stable input observers}\label{eq:subsec:observer}

In this subsection, on the basis of a.s.-left invertibility of the linear quantum system \eqref{Kalmansys} studied in the preceding subsection, we construct input observers which, as implied by Definition \ref{def:invertibility},  asymptotically reconstruct the input from the output. Two types of  input observers are constructed.

For the first type, we follow the construction procedure proposed in \cite{DD2023}. We look at the ``$co$" subsystem by ignoring the other modes, which is
\begin{equation}\label{minireal}\begin{aligned}
\dot{\mbf{x}}_{co}(t)&=A_{co}\mbf{x}_{co}(t)+B_{co}\mbf{u}(t), \\
\mbf{y}(t)&=C_{co}\mbf{x}_{co}(t)+\mbf{u}(t).
\end{aligned}\end{equation}
Notice that the $D$-matrix is an identity matrix. By the procedure given in \cite[Subsec. IV-C]{DD2023}, the input  observer is
\begin{equation}\label{eq:observer}\begin{aligned}
\dot{\xi}^o&=A_\ell\xi^o+B_{co}\mbf{y}, \\
u_\ell&=-C_{co}\xi^o+\mbf{y},
\end{aligned}\end{equation}
where $A_\ell=A_{co}-B_{co}C_{co}$.

\bmrk
In the input observer constructed in  \cite[Eq. (37)]{DD2023},  there is a feedback gain matrix $K_\ell$, and the resulting observer  is of the Luenberger type, where  the matrix $K_\ell$ guarantees the asymptotic stability of the input observer. For the linear quantum system \eqref{Kalmansys}, as the number of the outputs is equal to the number of the inputs (Remark \ref{rem:full_rank}), the input observer \eqref{eq:observer} need no such matrix $K_\ell$. In Theorem \ref{thm:stable_observer} below we show  that due to the special nature of the invariant zeros of linear quantum systems, the input observer \eqref{eq:observer} is  actually asymptotically stable.
\emrk

Define the state estimation error $\epsilon(t)=\xi^o(t)-\mbf{x}_{co}(t)$. Then we have
\begin{equation}
\dot{\epsilon}(t)=A_\ell \epsilon(t),
\end{equation}
and
\begin{equation}
\mbf{u}(t)-u_\ell(t)=C_{co}\epsilon(t).
\end{equation}

\begin{theorem} \label{thm:stable_observer}
If the linear quantum system \eqref{Kalmansys} is a minimal realization \eqref{minireal} and a.s.-left invertible, then it has an asymptotically stable input observer  \eqref{eq:observer}.
\end{theorem}

\emph{Proof.} As the linear quantum system \eqref{minireal} is a.s.-left invertible,  by Corollary \ref{cor:minimal} the set of eigenvalues of $A_{co}$ is in the right-half plane. In this case, by Theorem \ref{cor:invariant_0} all the invariant zeros of the linear quantum system are in the left-half plane. Since
\begin{equation*}
\left[\begin{array}{cc}
A_{co}-sI & B_{co} \\
C_{co} & I
\end{array}\right]\left[\begin{array}{cc}
I & 0 \\
-C_{co} & I
\end{array}\right]   =
\left[\begin{array}{cc}
A_\ell-sI & B_{co} \\
0 & I
\end{array}\right],
\end{equation*}
invariant zeros of the linear quantum system are exactly the eigenvalues of $A_\ell$. Thus, $\lambda(A_\ell)$ is in the left-half plane, and the observer \eqref{eq:observer} is asymptotically stable. \hfill $\blacksquare$

\bmrk
The stable input observer \eqref{eq:observer} may not be a valid quantum-mechanical system as it may not satisfy the physical realizability conditions. Nevertheless, it is indeed a stable input observer if we focus on the \emph{average} dynamics such as those given in Eq. \eqref{eq:real_sys_mean}. In other words, replacing $\mbf{x},  \mbf{y}, \mbf{u}$ by their mean values $\langle \mbf{x} \rangle, \langle \mbf{y} \rangle,\langle \mbf{u} \rangle $, then we have $\langle\mbf{u}(t)\rangle - u_\ell(t)  \to 0$ exponentially as $t\to \infty$.  In this case, it is a classical stable input observer for a linear classical system. However, due to the absence of the feedback gain matrix $K_\ell$, its form is still different from the one in \cite[Eq. (37)]{DD2023} (see also \cite{XS03}). Actually, it is the pole-zero duality of linear quantum systems that guarantees the stability of the input observer \eqref{eq:observer}.
\emrk

In contrast to the stable input observer constructed above using the procedure in \cite{DD2023}, in the following  we design another type of input observers for a.s.-left invertible linear quantum systems. Denote
\begin{equation}\label{T^prime}
\hspace{-1ex} \bar{J}_n = \left[
\begin{array}{ccc}
J_{n_{3}} & 0 & 0 \\
0 & J_{n_{1}} & 0 \\
0 & 0 & J_{n_{2}}%
\end{array}%
\right], \widetilde{J}_n= \left[\begin{array}{cc}
0 & I_n \\
I_n & 0
\end{array}\right],  T^\prime = \hat{T}^\dagger\widetilde{J}_n\hat{T},
\end{equation}
where $\hat{T}$ is the coordinates transformation in Eq. \eqref{eq:observables}.

\bthm \label{thm:stable_observer_2}
Suppose that the intrinsic system Hamiltonian $\mathbb{H}=0$ for the linear quantum system realization \eqref{Kalmansys}. A linear quantum system $W$ can be constructed as
\begin{equation}\label{sys_kal_inverse}\begin{aligned}
\dot{\mbf{x}}^\prime(t)&=\bar{A}^\prime \mbf{x}^\prime(t)+\bar{B}^\prime \mbf{y}(t), \\
\mbf{u}^\prime(t)&=\bar{C}^\prime \mbf{x}^\prime(t)+\mbf{y}(t),
\end{aligned}\end{equation}
where $\mbf{y}$ is the output of the system \eqref{Kalmansys}, and the real system matrices are
\begin{equation} \label{eq:ABCD_inverse_2}\begin{aligned}
\bar{C}^\prime=\bar{C}T^\prime, \ \bar{B}^\prime=\mathbb{\overline{J}}_n\bar{C}^{\prime\top}\mathbb{J}_m, \ \bar{A}^\prime=\frac{1}{2}\bar{B}^\prime\bar{C}^\prime.
\end{aligned}\end{equation}
$W$ is the inverse of the system \eqref{Kalmansys} in the sense that the transfer matrix $\mathbb{W}$ of the linear quantum system $W$ satisfies
\begin{equation}\label{eq:july6_PG}
\mathbb{W}(s)\mathbb{G}(s) = I.
\end{equation}
Moreover, if the linear quantum system \eqref{Kalmansys} is a minimal realization and a.s.-left invertible, then  the input observer  \eqref{sys_kal_inverse} is Hurwitz stable.
\ethm
\emph{Proof.} By means of the quantum system inversion techniques in \cite{GJN10}, we construct the linear quantum system $W$ with annihilation operators $\mbf{a}^\prime$. In the $(S,\mbf{L},\mbf{H})$ formalism, the scattering operator $S^\prime=I$, the intrinsic Hamiltonian $\mbf{H}^\prime=0$, and the  coupling operator is
\begin{equation}
\mbf{L}^\prime= \left[\begin{array}{cc}
C_+ & C_-
\end{array}\right]\breve{\mbf{a}}^\prime,
\end{equation}
where $C_-,C_+$ are the system parameters of the linear quantum system \eqref{Kalmansys}. According to the modeling described in Subsection \ref{subsec:model}, the linear quantum  system $W$ in the annihilation-creation operator form is
\begin{equation}\label{sys_kal_inverse_complex}\begin{aligned}
\dot{\breve{\mbf{a}}}^\prime (t)&=\mathcal{A}^\prime  \breve{\mbf{a}}^\prime (t)+\mathcal{B}^\prime \breve{\mbf{b}}_{\rm out}(t), \\
\tilde{\mbf{b}}^\prime(t)&=\mathcal{C}^\prime \breve{\mbf{a}}^\prime  +\breve{\mbf{b}}_{\rm out}(t),
\end{aligned}\end{equation}
where the system matrices
\begin{equation}
\mathcal{C}^\prime=\mathcal{C}\widetilde{J}_n,  \  \
\mathcal{B}^\prime=-\mathcal{C}^{\prime\flat}, \ \
\mathcal{A}^\prime=-\frac{1}{2}\mathcal{C}^{\prime\flat}\mathcal{C}^\prime.
\end{equation}
Applying the coordinates transformation
\begin{equation}\label{eq:transform}
\mbf{x}^\prime =\hat{T}^\dagger \breve{\mbf{a}}^\prime,
\ \mbf{u}^\prime =V_m\tilde{\mbf{b}}^\prime(t) , ~ \mbf{y}^\prime =V_m\breve{\mbf{b}}_{\rm out}(t),
\end{equation}
to Eq. \eqref{sys_kal_inverse_complex} yields  system \eqref{sys_kal_inverse} with system matrices
\begin{equation}\label{eq:ABCD_inverse}\begin{aligned}
\bar{C}^\prime
&=V_m\mathcal{C}^\prime \hat{T}= V_m V_m^\dagger\bar{C}\hat{T}^\dagger\widetilde{J}_n \hat{T}=\bar{C}\hat{T}^\dagger\widetilde{J}_n\hat{T} =\bar{C}T^\prime, \\
\bar{B}^\prime&=\hat{T}^\dagger\mathcal{B}^\prime V_m^\dagger=\bar{\mathbb{J}}_n\bar{C}^{\prime\top}\mathbb{J}_m, \\
\bar{A}^\prime&=\hat{T}^\dagger\mathcal{A}^\prime\hat{T}=\frac{1}{2}\bar{B}^\prime\bar{C}^\prime.
\end{aligned}\end{equation}
(Note that the relations $\hat{T}^\dagger J_n \hat{T}=\imath\bar{\mathbb{J}}_n$ and $V_m J_m V_m^\dagger=\imath\mathbb{J}_m$ have been used in the above derivation.)  Clearly, Eq. \eqref{eq:ABCD_inverse} is Eq. \eqref{eq:ABCD_inverse_2}.  Moreover, according to the quantum system inversion techniques \cite{GJN10}, the transfer matrix $\mathbb{W}(s)$ of the  input observer \eqref{sys_kal_inverse} satisfies Eq. \eqref{eq:july6_PG}. Next we show that  the system matrices $\bar{A}^\prime, \bar{B}^\prime, \bar{C}^\prime$ in Eq. \eqref{eq:ABCD_inverse_2} are all real-valued. Clearly, by  Eq. \eqref{eq:ABCD_inverse_2} it suffices to show that $T^\prime$ in Eq. \eqref{T^prime} is real.     As given by the first equation in the proof of Lemma 4.8 in \cite{ZGPG18}, $\hat{T}=T\tilde{V}_n^\dagger$, where the matrix $T$ is given in   \cite[Eq. (47)]{ZGPG18} and $\tilde{V}_n^\dagger$ is given in  \cite[Lem. 4.8]{ZGPG18}. Algebraic manipulations yield
\begin{equation} \label{eq:T_prime_inverse}
T^\prime=\left[\begin{array}{ccc}
\Pi\Lambda_{33}\Pi^\top  & \Pi\Lambda_{31} & \Pi\Lambda_{32} \\
 (\Pi\Lambda_{31})^\top  & \Lambda_{11} & \Lambda_{12} \\
(\Pi\Lambda_{32})^\top & \Lambda_{12}^\top & \Lambda_{22}
\end{array}\right],
\end{equation}
where, as given in \cite[Lem. 4.7]{ZGPG18},  the orthogonal and symplectic matrix
\begin{equation}\label{Pi}
\Pi=\left[\begin{array}{cccc}
I_{n_a} & 0 & 0 & 0 \\
0 & 0 & 0 & -I_{n_b} \\
0 & 0 & I_{n_a} & 0 \\
0 & I_{n_b} & 0 & 0
\end{array}\right],  \  n_a+n_b=n_3,
\end{equation}
and
\begin{equation}\label{Lambda}
\Lambda_{ij}=\left[\begin{array}{cc}
{\rm Re}(Z_i^\top Z_j) & -{\rm Im}(Z_i^\top Z_j) \\
-{\rm Im}(Z_i^\top Z_j) & -{\rm Re}(Z_i^\top Z_j)
\end{array}\right], \ i,j=1,2,3,
\end{equation}
with $Z_i$ being the unitary and Bogoliubov transformation matrices \cite[Thm. 4.1]{ZGPG18}.
Since $\hat{T}$ is unitary, $T^\prime$ is unitary too. By Eqs. \eqref{eq:T_prime_inverse} and \eqref{Lambda}, $T^\prime$ is real. As a result, $T'$ is both symmetric and orthogonal.  Therefore   the system matrices $\bar{A}^\prime, \bar{B}^\prime, \bar{C}^\prime$ in Eq. \eqref{eq:ABCD_inverse_2} are all real matrices. Moreover, noticing
\begin{equation*}\begin{aligned}
&T^{\prime}\mathbb{\overline{J}}_{n} T^\prime=
\hat{T}^\dagger\widetilde{J}_n\hat{T}\mathbb{\overline{J}}_{n} \hat{T}^\dagger\widetilde{J}_n\hat{T} =\tilde{V}_nT^\dagger\widetilde{J}_nT\tilde{V}_n^\dagger\mathbb{\overline{J}}_{n} \tilde{V}_nT^\dagger\widetilde{J}_nT\tilde{V}_n^\dagger \\
=&-\imath\tilde{V}_nT^\dagger\widetilde{J}_nT\bar{J}_nT^\dagger\widetilde{J}_nT\tilde{V}_n^\dagger =-\imath\tilde{V}_nT^\dagger\widetilde{J}_n J_n\widetilde{J}_nT\tilde{V}_n^\dagger \\
=&\imath\tilde{V}_nT^\dagger J_nT\tilde{V}_n^\dagger =\imath\tilde{V}_n\bar{J}_n\tilde{V}_n^\dagger
=-\bar{\mathbb{J}}_n,
\end{aligned}\end{equation*}
we have $\bar{\mathbb{J}}_nT^\prime\bar{\mathbb{J}}_n=T^\prime$. Hence,
\begin{equation}\begin{aligned}
		\bar{A}^\prime =\frac{1}{2}\bar{B}^\prime\bar{C}^\prime =-\mathbb{\overline{J}}_nT^\prime\mathbb{\overline{J}}_n\bar{A}T^\prime=-T^\prime\bar{A}T^\prime.
\end{aligned}\end{equation}
If the linear quantum system \eqref{Kalmansys} is a minimal realization and a.s.-left invertible, then by Corollary \ref{cor:minimal} $\lambda(\bar{A})$ is in the right-half plane, therefore the input observer \eqref{sys_kal_inverse} is Hurwitz stable.
Finally,  define $\varepsilon(t)=T^\prime\mbf{x}^\prime(t)+\mbf{x}(t)$. Then we have $\dot{\varepsilon}(t)=-\bar{A}\varepsilon(t)$, and the input estimation error $\mbf{u}^\prime(t)-\mbf{u}(t)=\bar{C}\varepsilon(t)$, which  asymptotically converges to $0$.  That is,  the input observer \eqref{sys_kal_inverse} is a stable input estimator. \hfill $\blacksquare$

\bmrk
In Theorem \ref{thm:stable_observer_2} it is assumed that $\mbb{H}=0$. This is not a very restrictive condition. When the quantum system of interest is resonant with the input, then $\mbb{H}=0$ in the rotating frame, see e.g., \cite[Section 1.5.1]{NY17}.
\emrk

\begin{example}
Let the system matrices of a linear quantum system in the annihilation-creation form \eqref{zerosys} be
$\mathcal{A}=\frac{\kappa}{2}I_2$, $\mathcal{B}=\mathcal{C}=\imath\sqrt{\kappa}\mathbb{J}_1$, and accordingly in the Kalman canonical form \eqref{Kalmansys} we have $\bar{A}=\frac{\kappa}{2}I_2$, $\bar{B}=\bar{C}=\sqrt{\kappa}\widetilde{J}_1$, the transfer matrix is $\mathbb{G}(s)=\frac{s+\frac{\kappa}{2}}{s-\frac{\kappa}{2}}I_2$. By Theorem \ref{thm:stable_observer_2}, the system matrices of input observer $W$ are designed as $\bar{A}^\prime=-\frac{\kappa}{2}I_2$, $\bar{B}^\prime=\bar{C}^\prime=-\sqrt{\kappa}\mathbb{J}_1$, and the corresponding transfer matrix $\mathbb{W}(s)=\frac{s-\frac{\kappa}{2}}{s+\frac{\kappa}{2}}I_2$. Clearly, this input observer is Hurwitz stable and Eq. \eqref{eq:july6_PG} holds.
\end{example}

\subsection{Strong left invertibility}\label{eq:subsec:Strong}

The following result is about the strong left invertibility of linear quantum  systems.

\begin{theorem}\label{s.-left inv.}
The linear quantum system \eqref{Kalmansys} is s.-left invertible if and only if $\lambda(-\bar{A}_c)\subseteq\lambda(\bar{A}_{\bar{o}})$.
\end{theorem}
\emph{Proof.} By \cite[Thm. 12]{DD2023} and Eq. \eqref{eq:normrank_P}, the linear quantum system \eqref{Kalmansys} is s.-left invertible if and only if the set of invariant zeros equals the set of output-decoupling zeros, which by Theorem \ref{cor:invariant_0} can be expressed as $\lambda(-\bar{A}_o)\cup\lambda(-\bar{A}_{\bar{o}})=\lambda(\bar{A}_{\bar{o}})$, or equivalently, $\lambda(-A_{co})\cup\lambda(-A_h^{22})\cup\lambda(-A_h^{11})\cup\lambda(-A_{\bar{c}\bar{o}})=\lambda(A_h^{11})\cup\lambda(A_{\bar{c}\bar{o}})$. By \cite[Lem. 3.1]{ZPL20}, we have $\lambda(-A_h^{22})=\lambda(A_h^{11})$, $\lambda(-A_{\bar{c}\bar{o}})=\lambda(A_{\bar{c}\bar{o}})$, and thus the necessary and sufficient condition reduces to $\lambda(-A_{co})\cup\lambda(-A_h^{11})\subseteq\lambda(A_h^{11})\cup\lambda(A_{\bar{c}\bar{o}})$, which is exactly $\lambda(-\bar{A}_c)\subseteq\lambda(\bar{A}_{\bar{o}})$.    \hfill $\blacksquare$

\bprop\la{eq:prop:s.-left invertible}
If  the linear quantum system \eqref{Kalmansys} has no the ``$co$" subsystem and  Assumption \ref{Feb-h} holds, then it is  s.-left invertible. Moreover, $\bar{C}=\bar{B}=0$, and thus there is only the ``$\bar{c}\bar{o}$" subsystem which is decoupled from the input-output channels.
\eprop
\emph{Proof.} Under Assumption \ref{Feb-h}, the necessary and sufficient condition in Theorem \ref{s.-left inv.} becomes $\lambda(-A_{co})\subseteq\lambda(\bar{A}_{\bar{o}})$. Clearly, if the linear quantum system \eqref{Kalmansys} has no the ``$co$" subsystem, the condition naturally holds and  it is  s.-left invertible.  Moreover, as  the linear quantum system \eqref{Kalmansys} has no the ``$co$" subsystem,    its transfer matrix $\mathbb{G}(s) \equiv  I$. Then for an arbitrary initial condition $\mbf{x}(0)=x_0$, we have $\mbf{y}(t) =\bar{C} e^{\bar{A}t}x_0 + \mbf{u}(t)$.
If $\mbf{u}(t) = -\bar{C} e^{\bar{A}t}x_0 $, then $\mbf{y}(t)=0 $ for all $t>0$.    However,  as the system is s. left invertible,  $\mbf{y}(t)=0 \Rightarrow \mbf{u}(t)=0$ for all $t>0$. As $x_0$ is arbitrary, we must have $\bar{C}=0$. By Eq. \eqref{eq:july5_ABCD} we get $\bar{B}=0$. The result follows. \hfill $\blacksquare$

\bmrk
By \cite[Thm. 1.8]{Hautus83}, a linear system is strongly observable (\cite[Def. 21]{DD2023}) if and only if it has no   invariant zeros. In the linear quantum realm, the non-existence of invariant zeros is equivalent to  the non-existence of eigenvalues of the $A$-matrix. Thus, it means that there is only the $D$-matrix.   In this case, the conditions of Proposition \ref{eq:prop:s.-left invertible} hold naturally. Consequently, the system is  s.-left invertible. Thus, in the quantum regime, strong observability is stronger than s.-left invertibility. The same is true in the classical regime; see \cite[Thm. 12]{DD2023} and \cite[Thm.  1.8]{Hautus83} provided that  the Rosenbrock system matrix is of full column rank. Finally by Theorem \ref{cor:invariant_0}, if the poles of the linear quantum system  \eqref{Kalmansys}   are all in the right-half plane,  it is strong* detectable (\cite[Def. 1.3]{Hautus83}) and thus has a strong observer \cite[Thms. 1.5, 1.6 and 1.12]{Hautus83}.
\emrk


\section{Fundamental tradeoffs between quantum squeezing and robustness}\label{TSQS}

Heisenberg's uncertainty principle establishes a lower bound for the product of standard deviations of quantum canonical conjugate operators (e.g., position and  momentum), thus imposing intrinsic uncertainty upon quantum systems. However, ``squeezing" allows suppressing uncertainty in one operator by amplifying its conjugate counterpart’s fluctuations. First experimentally demonstrated by Slusher \emph{et al.} \cite{Slusher1985}, squeezing reduces noise to enhance signal-to-noise ratios (SNRs), making squeezed light critical for quantum communication and teleportation. Driven by its broad applications, research on generating and enhancing quantum squeezing has flourished for decades. Notable advancements include measurement-based feedback  \cite{WM10}, and coherent feedback \cite{NaokiAkira2012,BZL12,BCMH24}.

\begin{figure}[thpb]
 \centering
 \includegraphics[scale=0.4]{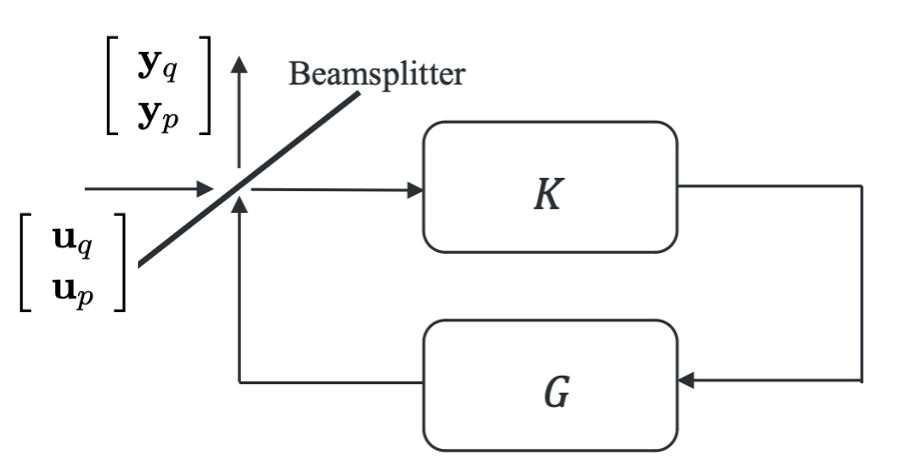}
 \caption{A quantum coherent feedback network composed of a linear quantum plant $G$,  a linear quantum controller $K$, and a beamsplitter.}
 \label{fig_LFT}
\end{figure}

In this section, we study another consequence of the pole-zero duality of linear quantum  systems explored in Section \ref{ZPLQS}, by demonstrating tradeoffs between quantum squeezing and sensitivity of the coherent feedback network, as shown in Fig. \ref{fig_LFT}.

\subsection{The coherent feedback network}\label{subsec:the network}

In this subsection, we describe the coherent feedback network as shown in Fig. \ref{fig_LFT}.

Firstly, we look at the quantum system $G$.  For simplicity, assume $G$ consists of a quantum harmonic oscillator ($n=1$) which is driven by a boson field ($m=1$), i.e.,  $G$ is a SISO system. (However, note that as a quantum field has both annihilation and creation operators, the size of the transfer matrix of the SISO system $G$ is actually $2\times 2$.)  In the $(S,\mbf{L},\mbf{H})$ formalism described in Subsection \ref{subsec:model}, assume that $\Omega$ is purely imaginary and $[C_- \ C_+]$ is  real or purely imaginary. Then the transfer  matrix of the quantum plant $G$ in terms of the Kalman canonical form \eqref{Kalmansys} is
\begin{equation} \label{eq:G__jun7}
\begin{aligned}
 &\mathbb{G}(s)=
\left[\begin{array}{cc}
\mathbb{G}_q(s) & 0 \\
0 & \mathbb{G}_p(s)
\end{array}\right]\\
=&
\left[\begin{array}{cc}
\frac{s+\imath\Omega_+-\frac{1}{2}\mathbb{C}_q\mathbb{C}_p}{s+\imath\Omega_++\frac{1}{2}\mathbb{C}_q\mathbb{C}_p} & 0 \\
0 & \frac{s-\imath\Omega_+-\frac{1}{2}\mathbb{C}_q\mathbb{C}_p}{s-\imath\Omega_++\frac{1}{2}\mathbb{C}_q\mathbb{C}_p}
\end{array}\right],
\end{aligned}\end{equation}
where $\mathbb{C}_q=C_-+C_+$ and $\mathbb{C}_p=C_--C_+$. Note that $\Omega_-=0$ as $\Omega$ is Hermitian. Clearly, the transfer matrix $\mathbb{G}(s)$ satisfies
\begin{equation}\label{tradeoffG}
\mathbb{G}_q(s)\mathbb{G}_p(-s)=1.
\end{equation}
(See also \cite[Eq. (11)]{Yan07}.)

Secondly, let $\Omega^\prime$ and $[C_-' \ C_+']$ be the parameters of the SISO quantum controller $K$, where $\Omega^\prime$ is purely imaginary and $\left[\begin{array}{cc}
C_-^\prime & C_+^\prime \\
\end{array}\right]$ is real or purely imaginary. Denote $\mathbb{C}_q^\prime=C_-^\prime+C_+^\prime$ and $\mathbb{C}_p^\prime=C_-^\prime-C_+^\prime$. Clearly, the corresponding transfer matrix $\mathbb{K}(s)$ is of the form as $\mathbb{G}(s)$ in Eq. \eqref{eq:G__jun7}.

Finally, let the beamsplitter be  $\left[\begin{smallmatrix}
\alpha & \beta \\
\beta & -\alpha
\end{smallmatrix}\right]$
with real parameters $\alpha$, $\beta$ satisfying $\alpha^2+\beta^2=1$.  Then it can be easily shown that  the closed-loop transfer matrix from $\mathbf{u}=\left[\begin{smallmatrix}
\mathbf{u}_q \\
\mathbf{u}_p
\end{smallmatrix}\right]$ to $\mathbf{y}=\left[\begin{smallmatrix}
\mathbf{y}_q \\
\mathbf{y}_p
\end{smallmatrix}\right]$ of the coherent feedback network in Fig. \ref{fig_LFT} is

\begin{equation}\label{Mar15-6}
\begin{aligned}
&\mathbb{T}(s) = (I+\alpha \mathbb{G}(s)\mathbb{K}(s))^{-1}(\alpha I+\mathbb{G}(s)\mathbb{K}(s))
\\
=&
\left[\begin{array}{cc}
\mathbb{T}_q(s) & 0 \\
0 & \mathbb{T}_p(s)
\end{array}\right]
= \left[
\bey{cc}
\frac{\alpha+\mathbb{G}_q(s)\mathbb{K}_q(s)}{1+\alpha\mathbb{G}_q(s)\mathbb{K}_q(s)} &0 \\
0 & \frac{\alpha+\mathbb{G}_p(s)\mathbb{K}_p(s)}{1+\alpha\mathbb{G}_p(s)\mathbb{K}_p(s)}
\eey
\right].
\end{aligned}
\end{equation}

It can be verified that $\mathbb{T}_q(s)\mathbb{T}_p(-s)=1$, which means that the structure in Eq. \eqref{tradeoffG} is preserved in this coherent feedback network; see also \cite{Yan07}. Also, it can be seen that $s_0$ is a pole of $\mathbb{T}(s)$ if and only if $-s_0$ is a transmission zero of $\mathbb{T}(s)$, which confirms Theorem \ref{cor:s_-s}.


\subsection{Ideal squeezing}\label{subsec:squeezing}

In this subsection, we study how to realize ideal squeezing using the coherent feedback network in Fig. \ref{fig_LFT}.

The squeezing ratio of a quantum input-output system is given by the ratio between the variance of the output quadrature and that of the corresponding input quadrature. If the initial joint system-field state is the vacuum state, then the squeezing ratio of the amplitude quadrature of the coherent feedback network in Fig. \ref{fig_LFT} at the frequency $\omega$ is given by $\frac{\langle \mathbf{y}_q^2(\imath \omega) \rangle}{\langle \mathbf{u}_q^2(\imath \omega) \rangle}$, and similarly the squeezing ratio of the phase quadrature at the frequency $\omega$  is given by $\frac{\langle \mathbf{y}_p^2(\imath \omega) \rangle}{\langle \mathbf{u}_p^2(\imath \omega) \rangle}$. A zero squeezing ratio is often referred to as \emph{ideal squeezing} or \emph{infinite squeezing}.

The following result shows that ideal squeezing is achieved at  zeros of a transfer function.
\bthm \label{thm:blocking_zero_squeezing}
$\imath \omega $ is a  zero of the transfer function $\mathbb{T}_q(s)$ in Eq. \eqref{Mar15-6} if and only if at which ideal squeezing in the $q$ quadrature is achieved.
\ethm
\emph{Proof.} Let $\imath \omega $ be a  zero of the transfer function $\mathbb{T}_q(s)$ in Eq. \eqref{Mar15-6}. Then   $\mathbb{T}_q(\imath \omega)=0$ and thus $\mathbf{y}_q(\imath \omega)=0$ and ideal squeezing in the $q$ quadrature is achieved. The above procedure  can be reversed. \hfill $\blacksquare$

On the basis of Theorem \ref{thm:blocking_zero_squeezing}, the following result shows when ideal squeezing at zero frequency can be realized by the coherent feedback network in Fig. \ref{fig_LFT}.
\begin{theorem}\label{thm:squeezing_alpha}
If \begin{equation}\label{Mar15-3}\begin{aligned}
&(1+\alpha)\left(\frac{1}{4}\mathbb{C}_q\mathbb{C}_p\mathbb{C}_q^\prime\mathbb{C}_p^\prime-\Omega_+\Omega_+^\prime\right) \\
-\ &(1-\alpha)\left(\frac{\imath}{2}\mathbb{C}_q\mathbb{C}_p\Omega_+^\prime+\frac{\imath}{2}\mathbb{C}_q^\prime\mathbb{C}_p^\prime\Omega_+\right)=0,
\end{aligned}\end{equation}
then $\mathbb{T}_q(0)=0$, and thus the coherent feedback network in Fig. \ref{fig_LFT} achieves ideal squeezing in the $q$ quadrature at zero frequency, namely $\frac{\langle \mathbf{y}_q^2(0) \rangle}{\langle \mathbf{u}_q^2(0) \rangle}=0$. Similarly, if \begin{equation}\label{Mar15-4}\begin{aligned}
&(1+\alpha)\left(\frac{1}{4}\mathbb{C}_q\mathbb{C}_p\mathbb{C}_q^\prime\mathbb{C}_p^\prime-\Omega_+\Omega_+^\prime\right) \\
+\ &(1-\alpha)\left(\frac{\imath}{2}\mathbb{C}_q\mathbb{C}_p\Omega_+^\prime+\frac{\imath}{2}\mathbb{C}_q^\prime\mathbb{C}_p^\prime\Omega_+\right)=0,
\end{aligned}\end{equation}
then $\mathbb{T}_p(0)=0$, and therefore  the coherent feedback network in Fig. \ref{fig_LFT} achieves ideal squeezing in the $p$ quadrature at zero frequency, namely $\frac{\langle \mathbf{y}_p^2(0) \rangle}{\langle \mathbf{u}_p^2(0) \rangle}=0$.
\end{theorem}
\emph{Proof.}
By Eq. \eqref{Mar15-6}, to realize ideal squeezing of $\frac{\langle \mathbf{y}_q^2(\imath \omega) \rangle}{\langle \mathbf{u}_q^2(\imath \omega) \rangle}$ at the frequency $\omega$, we need $\mathbb{T}_q(\imath \omega)=0$, which means that either $\alpha+\mathbb{G}_q (\imath \omega)\mathbb{K}_q(\imath \omega)= 0$ or $|1+\alpha\mathbb{G}_q(\imath \omega)\mathbb{K}_q(\imath \omega)| = \infty$. However, $|1+\alpha\mathbb{G}_q(\imath \omega)\mathbb{K}_q(\imath \omega)|= \infty$ is equivalent to $|\mathbb{G}_j(\imath \omega)\mathbb{K}_q(\imath \omega)|= \infty$ and $\alpha\neq0$, which means $\mathbb{T}_q(\imath \omega)=\frac{1}{\alpha} \neq 0$. Thus, ideal squeezing in the $q$ quadrature requires that $\alpha+\mathbb{G}_q(\imath \omega)\mathbb{K}_q(\imath \omega)= 0$, or equivalently,
\begin{equation}\label{Mar15-2}\begin{aligned}
\alpha+\frac{\imath \omega+\imath\Omega_+-\frac{1}{2}\mathbb{C}_q\mathbb{C}_p}{\imath \omega+\imath\Omega_++\frac{1}{2}\mathbb{C}_q\mathbb{C}_p}\times     \frac{\imath \omega+\imath\Omega_+^\prime-\frac{1}{2}\mathbb{C}_q^\prime\mathbb{C}_p^\prime}{\imath \omega+\imath\Omega_+^\prime+\frac{1}{2}\mathbb{C}_q^\prime\mathbb{C}_p^\prime}=0,
\end{aligned}\end{equation}
which can be rewritten as  Eq. \eqref{Mar15-3} when $\omega=0$. Hence $\mathbb{T}_q(0)=0$ if Eq. \eqref{Mar15-3} holds. Hence, by Theorem \ref{thm:blocking_zero_squeezing}  $\frac{\langle \mathbf{y}_q^2(0) \rangle}{\langle \mathbf{u}_q^2(0) \rangle}=0$.  The $p$ quadrature case can be proved similarly. \hfill $\blacksquare$

The following two examples demonstrate that the coherent feedback network in Fig. \ref{fig_LFT} can realize ideal squeezing at zero frequency by means of a passive  or an active controller.

\begin{example}\label{example_bs}
Assume that the controller $K=1$. Thus the coherent feedback network in Fig. \ref{fig_LFT} consists of the system $G$ and a beamsplitter.
To realize ideal squeezing of the $q$ quadrature at zero frequency, in other words, $\mathbb{T}_q(0)=0$, by Eq. \eqref{Mar15-3} we require that $\frac{1-\alpha}{2}\mathbb{C}_q\mathbb{C}_p-\imath(1+\alpha)\Omega_+=0$,
which implies that $\alpha=-\frac{\imath\Omega_+-\frac{1}{2}\mathbb{C}_q\mathbb{C}_p}{\imath\Omega_++\frac{1}{2}\mathbb{C}_q\mathbb{C}_p}$. Interestingly, by comparing Eqs. \eqref{Mar15-3} and \eqref{Mar15-4} the ideal squeezing of the $p$ quadrature at zero frequency can be obtained by setting $\alpha  \to \frac{1}{\alpha}$.
\end{example}

\begin{example}\label{example_ac}
In this example, we design a dynamic controller $K$ instead of the static one in Example \ref{example_bs}. For simplicity, let $\mathbb{C}_q^\prime=\mathbb{C}_q$ and  $\mathbb{C}_p^\prime=\mathbb{C}_p$. Then the constraints in Eqs. \eqref{Mar15-3} and \eqref{Mar15-4} reduce to
\begin{equation}\label{Mar15-9}\begin{aligned}
(1+\alpha)\left(\frac{\mathbb{C}_q^2\mathbb{C}_p^2}{4}-\Omega_+\Omega_+^\prime\right)\mp\frac{\imath(1-\alpha)}{2}\mathbb{C}_q\mathbb{C}_p(\Omega_++\Omega_+^\prime)=0,
\end{aligned}\end{equation}
which yields
\begin{equation}\label{Mar15-10}
\Omega_+^\prime=\frac{\mp\imath\mathbb{C}_q\mathbb{C}_p}{2}
\frac{(1+\alpha)\mathbb{C}_q\mathbb{C}_p\mp2(1-\alpha)\imath\Omega_+}{(1-\alpha)\mathbb{C}_q\mathbb{C}_p\mp2(1+\alpha)\imath\Omega_+},
\end{equation}
where ``$-$" and ``$+$" signs correspond to the $q$ quadrature and $p$ quadrature, respectively.
With this controller $K$, the coherent feedback network in Fig. \ref{fig_LFT}  can realize ideal squeezing at zero frequency.
\end{example}

In Example \ref{example_bs}, the beamsplitter parameter $\alpha=-\frac{\imath\Omega_+-\frac{1}{2}\mathbb{C}_q\mathbb{C}_p}{\imath\Omega_++\frac{1}{2}\mathbb{C}_q\mathbb{C}_p}$ for ideal squeezing in the $q$ quadrature, and ideal squeezing in the $p$ quadrature is achieved by setting $\alpha \to 1/\alpha$. However, if the calculated value of $\alpha$ is not in the interval $[-1,1]$,  the   coherent feedback network cannot realize ideal squeezing at zero frequency. Thus in this case, an active controller proposed in Example \ref{example_ac} is needed to realize ideal squeezing, which is independent of $\alpha$.


\subsection{Squeezing and sensitivity}\label{subsec:squeezing_robustness}

In this subsection, we study the tradeoff between quantum squeezing and  robustness of the coherent feedback network  with respect to parameter variations in the quantum plant $G$ in Fig. \ref{fig_LFT}.

Define
\beq \la{eq:T_G_S}
\mathbb{S}_j \triangleq  \frac{d\log \mathbb{T}_j}{d\log \mathbb{G}_j}, ~~ j=q,p.
\eeq
Substituting Eqs. \eqref{eq:G__jun7} and \eqref{Mar15-6} into Eq. \eqref{eq:T_G_S} yields
\begin{equation} \la{eq:S}
\begin{aligned}
&\mathbb{S}=\left[\begin{array}{cc}
\mathbb{S}_q & 0 \\
0 & \mathbb{S}_p
\end{array}\right]
\\
=&\;
\left[\begin{array}{cc}
\frac{\beta^2\mathbb{G}_q\mathbb{K}_q}{(1+\alpha\mathbb{G}_q\mathbb{K}_q)(\alpha +\mathbb{G}_q\mathbb{K}_q)} & 0 \\
0 & \frac{\beta^2\mathbb{G}_p\mathbb{K}_p}{(1+\alpha\mathbb{G}_p\mathbb{K}_p)(\alpha +\mathbb{G}_p\mathbb{K}_p)}
\end{array}\right].
\end{aligned}
\end{equation}
By definition, $\mathbb{S}$ describes the sensitivity of the closed-loop input-output relation $\mathbb{T}$ with respect to the parameter variations in the quantum plant $G$. In this sense, we call $\mathbb{S}$ the \emph{sensitivity function}. In fact, this is the original definition of the sensitivity function in classical control theory; see, e.g., \cite[Ch. 4]{bode1945network}, \cite[Sect. 3.4]{DFT13}, \cite[Eq. (2.24)]{SP05}, \cite[Ch. 12]{AM21}.

If the coherent feedback network in Fig. \ref{fig_LFT}  is designed to realize ideal squeezing at some frequency $s_0$ for the input quadrature $j=q$ or $p$, then we need $\alpha+\mathbb{G}_j(s_0)\mathbb{K}_j(s_0) =0$ in Eq. \eqref{Mar15-6}, but by Eq. \eqref{eq:S}  this implies the corresponding sensitivity function $\mathbb{S}_j(s_0)=\infty$.  Thus, if the coherent feedback network is designed to realize ideal squeezing,  it will be extremely sensitive to the parameter variations in quantum plant $G$. This reveals a fundamental tradeoff between squeezing and system robustness posed by the zeros of the coherent feedback network in Fig. \ref{fig_LFT}.

Note that the transfer matrix $\mathbb{T}$ in Fig. \ref{fig_LFT} is not the \emph{complementary sensitivity function} commonly used in classical control theory (namely, the transfer function from the measurement noise to the system output, see, e.g., \cite[Fig. 1]{CI19} and \cite[Fig. 12.9]{AM21}), therefore the well-known \emph{complementarity constraint} $\mathbb{S}(s)+\mathbb{T}(s)\equiv 1$ no longer applies. Nevertheless, as $\mathbb{T}$ characterizes the input-output squeezing while  $\mathbb{S}$ characterizes the sensitivity of  $\mathbb{T}$ with respect to uncertainties in the quantum plant $G$, it is still meaningful to investigate $\mathbb{S}(s)+\mathbb{T}(s)$ as it reveals the tradeoff between squeezing and sensitivity. The following theorem is the main result of this subsection.

\begin{theorem} \label{thm:S+T}
For the coherent feedback network in Fig. \ref{fig_LFT}, $\mathbb{S}+\mathbb{T}$ can take any real values including $\pm
\infty$ by selecting appropriate controller and beamsplitter parameters.
\end{theorem}

\emph{Proof.}
Re-write the sensitivity matrix  $\mathbb{S}$ in Eq. \eqref{eq:S} as
\beq\label{eq:s_july4}
\mathbb{S}= \frac{\mathbb{GK}}{I-(\mathbb{GK})^2} \frac{1-\mathbb{T}^2}{\mathbb{T}}.
\eeq
According to Eqs. \eqref{Mar15-6} and \eqref{eq:s_july4},
\beq\la{eq:S+T}\begin{aligned}
\mathbb{S}+\mathbb{T} =&\frac{\alpha^2 I+ (1+2\alpha-\alpha^2)\mathbb{GK} + (\mathbb{GK})^2 }{(I+\alpha\mathbb{G}\mathbb{K})(\alpha I+\mathbb{G}\mathbb{K})}
\\
=&\frac{(\mathbb{GK} + \frac{1+2\alpha-\alpha^2}{2}I)^2 - \frac{(1+4\alpha-\alpha^2) (1-\alpha^2)}{4}I}{(I+\alpha\mathbb{G}\mathbb{K})(\alpha I+\mathbb{G}\mathbb{K})}.
\end{aligned}
\eeq

(Notice that $\mathbb{G}$, $\mathbb{K}$, $\mathbb{S}$, and $\mathbb{T}$ in Eqs. \eqref{eq:s_july4}-\eqref{eq:S+T} are all 2-by-2 diagonal matrices. Thus all these equations should be understood as diagonal matrix equations. For example,  Eq. \eqref{eq:S+T} is in fact ${\rm diag}\{(\mathbb{S}+\mathbb{T})_q, (\mathbb{S}+\mathbb{T})_p\}$, where
\beq \label{eq:may8_S+T_j}
(\mathbb{S}+\mathbb{T})_j
=
\frac{(\mathbb{G}_j\mathbb{K}_j + \frac{1+2\alpha-\alpha^2}{2})^2 - \frac{(1+4\alpha-\alpha^2) (1-\alpha^2)}{4}}{(1+\alpha\mathbb{G}_j\mathbb{K}_j)(\alpha+\mathbb{G}_j\mathbb{K}_j)}
\eeq
for $j=q,p$. This convention will be used in the following discussions for notational convenience.)  We focus on the $q$ quadrature case only as the $p$ quadrature case can be established similarly.

Given the quantum plant $G$,   we can always find  parameters $\mathbb{C}_q^\prime$, $\mathbb{C}_p^\prime$, and $\Omega_+^\prime$ of the controller $K$ such that
\begin{equation}\label{eq:may8_KG}
\begin{aligned}
\Omega_+\mathbb{C}_q\mathbb{C}_p+\Omega_+^\prime\mathbb{C}_q^\prime\mathbb{C}_p^\prime &=0, \\
\frac{1}{4}\mathbb{C}_q^2\mathbb{C}_p^2-\Omega_+^2 &=\frac{1}{4}\mathbb{C}_q^{\prime 2}\mathbb{C}_p^{\prime 2}-\Omega_+^{\prime 2}
\end{aligned}\end{equation}
hold, which means that $\mathbb{G}_q(s)\mathbb{K}_q(s)=\mathbb{G}_p(s)\mathbb{K}_p(s)$, and thus $(\mathbb{S}+\mathbb{T})_q=(\mathbb{S}+\mathbb{T})_p$, $\forall  s\in \mathbb{C}$. For the quantum controller $K$ satisfying Eq. \eqref{eq:may8_KG}, in what follows we show that for any given $\alpha \in [-1,1]$, there always exists some  $s_0\in \mathbb{C}$ such that
\beq \label{eq:GK-I}
\mathbb{G}_q(s_0)\mathbb{K}_q(s_0) = -\frac{1+2\alpha-\alpha^2}{2},
\eeq
and thus by Eq. \eqref{eq:may8_S+T_j},
\begin{equation}\label{June21-1}
\mathbb{S}_q(s_0)+\mathbb{T}_q(s_0) = \frac{1+4\alpha-\alpha^2}{(2-\alpha)(1-\alpha^2)}.
\end{equation}

For ease of representation, Eq. \eqref{eq:GK-I} can be re-written as
\begin{equation}\label{June18-5}
a_2s^2+a_1s+a_0=0,
\end{equation}
where the real-valued coefficients are
\begin{equation}\label{June18-4}\begin{aligned}
a_2=&\frac{(\alpha+1)(\alpha-3)}{2}, \\
a_1=&\frac{\imath(\alpha+1)(\alpha-3)}{2}(\Omega_++\Omega_+^\prime) \\
&+\frac{(\alpha-1)^2}{4}(\mathbb{C}_q\mathbb{C}_p+\mathbb{C}_q^\prime\mathbb{C}_p^\prime), \\
a_0=&-\frac{1+2\alpha-\alpha^2}{2}(\imath\Omega_++\frac{1}{2}\mathbb{C}_q\mathbb{C}_p)(\imath\Omega_+^\prime+\frac{1}{2}\mathbb{C}_q^\prime\mathbb{C}_p^\prime) \\
&-(\imath\Omega_+-\frac{1}{2}\mathbb{C}_q\mathbb{C}_p)(\imath\Omega_+^\prime-\frac{1}{2}\mathbb{C}_q^\prime\mathbb{C}_p^\prime).
\end{aligned}\end{equation}

For each $\alpha \in (-1,1)$,  the quadratic equation in Eq. \eqref{June18-5} has a solution $s_0$. Thus, the equation in  Eq. \eqref{June21-1} always has a solution $s_0$ for each $\alpha \in (-1,1)$.   Clearly, the right-hand side of Eq. \eqref{June21-1} is a continuous and monotonically increasing function of $\alpha\in(-1,1)$ and it approaches $\pm  \infty$ as $\alpha$ goes to  $\pm 1$.  Furthermore, $(\mathbb{S}+\mathbb{T})_j$ attains $\pm\infty$ when $\alpha=\pm1$, $j=q,p$.  \hfill $\blacksquare$

The following two examples  demonstrate Theorem \ref{thm:S+T}.

\begin{example} (Example \ref{example_bs} revisited)
This is the setup studied in \cite{GW09,BZL12}, where $G$ is a degenerate parametric amplifier (DPA)  and $K$ is 1. A model of a DPA in the annihilation-creation operator form is, \cite[pp. 220]{GZ00},
\begin{equation*}\begin{aligned}
\dot{\breve{\mbf{a}}} &=-\frac{1}{2}\left[
\begin{array}{cc}
\kappa  & -\epsilon  \\
-\epsilon  & \kappa
\end{array}%
\right] \breve{\mbf{a}}-\sqrt{\kappa}\ \breve{\mbf{b}}_{\rm in},
\\
\breve{\mbf{b}}_{\rm out} &= \sqrt{\kappa}\ \breve{\mbf{a}} + \breve{\mbf{b}}_{\rm in},
\end{aligned}\end{equation*}
which in the real-quadrature operator representation is
\begin{equation}\label{eq:DPA_Sept6}
\begin{aligned}
\dot{\mbf{x}}=&-\frac{1}{2}\left[
\begin{array}{cc}
\kappa-\epsilon  & 0 \\
0 & \kappa+\epsilon
\end{array}
\right] \mbf{x}-\sqrt{\kappa}\ \mbf{u}, \\
\mbf{y}=& \sqrt{\kappa}\ \mbf{x} + \mbf{u}.
\end{aligned}\end{equation}
For this system, $\Omega_-=0$, $\Omega_+=\frac{i\epsilon}{2}$, $C_-=\sqrt{\kappa}$, and $C_+=0$. The parameter $\epsilon$ in $\Omega_+$ designates the strength of the pump field on the DPA. The transfer matrix of $G$ is
\begin{equation}
\mathbb{G}(s)=
\left[\begin{array}{cc}
\frac{s-\frac{\epsilon}{2}-\frac{\kappa}{2}}{s-\frac{\epsilon}{2}+\frac{\kappa}{2}} & 0 \\
0 & \frac{s+\frac{\epsilon}{2}-\frac{\kappa}{2}}{s+\frac{\epsilon}{2}+\frac{\kappa}{2}}
\end{array}\right].
\end{equation}
The sensitivity matrix function $\mathbb{S}$ and the closed-loop transfer matrix $\mathbb{T}$ can be calculated as
{\small
\begin{equation*}\begin{aligned}
\mathbb{S}_j (s)
=\frac{\beta^2\left[(s\mp\frac{\epsilon}{2})^2-\frac{\kappa^2}{4}\right]}
{(1+\alpha^2)\left[(s\mp\frac{\epsilon}{2})^2-\frac{\kappa^2}{4}\right]+2\alpha\left[(s\mp\frac{\epsilon}{2})^2+\frac{\kappa^2}{4}\right]},
\end{aligned}\end{equation*}}
\begin{equation}\begin{aligned}
\mathbb{T}_j(s)=\frac{(1+\alpha)(s\mp\frac{\epsilon}{2})-(1-\alpha)\frac{\kappa}{2}}{(1+\alpha)(s\mp\frac{\epsilon}{2})+(1-\alpha)\frac{\kappa}{2}}, ~~ j=q,p,
\end{aligned}\end{equation}
respectively. Set
\begin{equation}\label{aug23-1}
\epsilon=\mp\frac{1-\alpha}{1+\alpha}\kappa.
\end{equation}
Then $\mathbb{T}_q(0)=0$, $\mathbb{S}_q(0)=\infty$, or $\mathbb{T}_p(0)=0$, $\mathbb{S}_p(0)=\infty$. Thus, the ideal squeezing of the coherent feedback network can be realized in the $q$ quadrature or $p$ quadrature at zero frequency, while  the coherent feedback network will be extremely sensitive to  parameter variations.
\end{example}

\begin{remark}
By the ideal squeezing realization condition \eqref{aug23-1}, in the  $p$ quadrature we have $\epsilon=\frac{1-\alpha}{1+\alpha}\kappa$. The beamsplitter parameter can be solved as $\alpha=\frac{\kappa-\epsilon}{\kappa+\epsilon}$, which is exactly the critical value $\alpha_{\rm crit}$ of the feedback-enhanced squeezing scenario given in \cite[Eq. (39)]{GW09}.
\end{remark}

\begin{example}\label{Ex.5.4} (Example \ref{example_ac} revisited)
Assume that the plant $G$ and the controller $K$ are two DPAs with parameters $\epsilon_i$, $\kappa_i$, $i=1,2$, in Eq. \eqref{eq:DPA_Sept6} respectively. Then the  closed-loop transfer matrix of the coherent feedback network is
\begin{equation}\label{May13-1}\begin{aligned}\resizebox{1\linewidth}{!}{\ensuremath{\displaystyle
\mathbb{T}_j(s)=\frac{(1+\alpha)\left[(s\mp\frac{\epsilon_1}{2})(s\mp\frac{\epsilon_2}{2})+\frac{\kappa_1\kappa_2}{4}\right]-(1-\alpha)\left[\frac{\kappa_1}{2}(s\mp\frac{\epsilon_2}{2})+\frac{\kappa_2}{2}(s\mp\frac{\epsilon_1}{2})\right]}{(1+\alpha)\left[(s\mp\frac{\epsilon_1}{2})(s\mp\frac{\epsilon_2}{2})+\frac{\kappa_1\kappa_2}{4}\right]+(1-\alpha)\left[\frac{\kappa_1}{2}(s\mp\frac{\epsilon_2}{2})+\frac{\kappa_2}{2}(s\mp\frac{\epsilon_1}{2})\right]}}},
\end{aligned}\end{equation}
and the sensitivity function matrix is given in Eq. \eqref{eq:aug21_temp1},
\begin{figure*}
\begin{align} \label{eq:aug21_temp1}
\mathbb{S}_j(s)=\frac{\beta^2\left[\left(s\mp\frac{\epsilon_1}{2}\right)^2-\frac{\kappa_1^2}{4}\right]\left[\left(s\mp\frac{\epsilon_2}{2}\right)^2-\frac{\kappa_2^2}{4}\right]}
{(1+\alpha^2)\left[\left(s\mp\frac{\epsilon_1}{2}\right)^2-\frac{\kappa_1^2}{4}\right]\left[\left(s\mp\frac{\epsilon_2}{2}\right)^2-\frac{\kappa_2^2}{4}\right]
+2\alpha\left\{\left[\left(s\mp\frac{\epsilon_1}{2}\right)^2+\frac{\kappa_1^2}{4}\right]\left[\left(s\mp\frac{\epsilon_2}{2}\right)^2+\frac{\kappa_2^2}{4}\right]+\left(s\mp\frac{\epsilon_1}{2}\right)\left(s\mp\frac{\epsilon_2}{2}\right)\kappa_1\kappa_2\right\}},
\end{align}
\end{figure*}
$j=q$, $p$. Set
\begin{equation}\label{aug23-2}
\kappa_1\epsilon_2+\kappa_2\epsilon_1=\mp\frac{1+\alpha}{1-\alpha}(\kappa_1\kappa_2+\epsilon_1\epsilon_2).
\end{equation}
Then $\mathbb{T}_q(0)=0$ and $\mathbb{S}_q(0)=\infty$, or $\mathbb{T}_p(0)=0$ and  $\mathbb{S}_p(0)=\infty$. Thus, the coherent feedback network achieves ideal squeezing in either the $q$ or the $p$ quadrature at zero frequency. Meanwhile, this configuration exhibits extreme sensitivity to parameter variations.
\end{example}

\begin{figure}[!t]
  \centering
  \subfloat[Zeros of $\mathbb{T}_q(s)$ (blue) and poles of $\mathbb{T}_p(s)$ (red).]{
    \includegraphics[width=0.45\columnwidth]{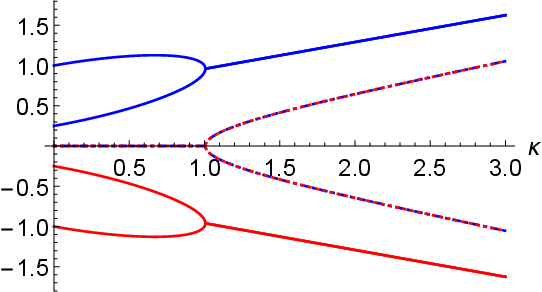}
  }
  \hfil 
  \subfloat[Zeros of $\mathbb{T}_p(s)$ (blue) and poles of $\mathbb{T}_q(s)$ (red).]{
    \includegraphics[width=0.45\columnwidth]{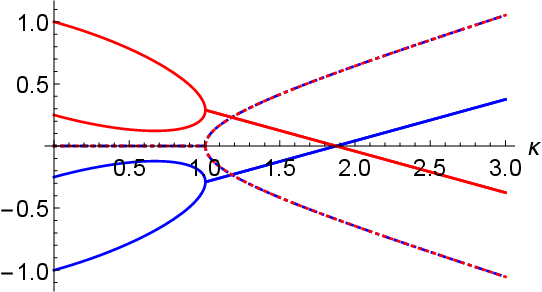}
  }
    \caption{Pole-zero duality in Example \ref{Ex.5.4}. The vertical axis denotes the values for zeros and poles. Specifically, (a) the real (resp. imaginary) parts of zeros of $\mathbb{T}_q(s)$ are plotted in blue solid (resp. dotted) lines, while  the real (resp. imaginary) parts of poles of $\mathbb{T}_p(s)$ are plotted in red solid (resp. dotted) lines. (b) The real (resp. imaginary) parts of zeros of $\mathbb{T}_p(s)$ are plotted in blue solid  (resp. dotted) lines, while  the real (resp. imaginary)  parts of poles of $\mathbb{T}_q(s)$ are plotted in red solid  (resp. dotted) lines. Moreover, it can be seen that the poles of both $\mathbb{T}_q(s)$ and $\mathbb{T}_p(s)$ are stable when $\kappa>2$.}
  \label{fig:total_figure}
\end{figure}

Finally, we set $\alpha=0.2$, $\epsilon_1=2$, $\epsilon_2=0.5$ and $\kappa_1=\kappa_2=\kappa$ in Eq. \eqref{May13-1}. Fig. \ref{fig:total_figure} depicts the zeros and poles of $\mathbb{T}_q(s)$ and $\mathbb{T}_p(s)$ as the parameter $\kappa$ varies. It can be easily seen that the zeros of $\mathbb{T}_q(s)$ and the poles of $\mathbb{T}_p(s)$ consistently demonstrate opposite signs in Fig. \ref{fig:total_figure} (a), similar observations for the poles of $\mathbb{T}_q(s)$ and the zeros of $\mathbb{T}_p(s)$ are given in Fig. \ref{fig:total_figure} (b).  Thus both cases  confirm the validity of $\mathbb{T}_q(s)\mathbb{T}_p(-s)=1$.

\section{Concluding remarks and discussions}\label{conclu}
In this paper, we investigated the zeros and poles of linear quantum systems. We proved that $s_0$ is a zero of a linear quantum  system if and only if $-s_0$ is a pole, which means  that a linear quantum system is necessarily non-minimum phase if it is Hurwitz stable. As two applications of such pole-zero duality,  we derived necessary and sufficient conditions for the strong left invertibility of linear quantum systems based on their invariant zeros and also constructed stable input observers. Moreover, we examined fundamental trade-offs between input-output squeezing and sensitivity for linear coherent feedback networks.  The following discussions point toward three possible future research directions.

\begin{itemize}

\item The proposed two types of stable input observers were constructed on the basis of the pole-zero correspondence for linear quantum systems, which  in general does  not hold for linear classical systems. In other words, their construction is based on quantum mechanics. However, they should be understood as input observers for average dynamics. Thus, an open question is how to construct a stable input observer for linear quantum systems instead of their average dynamics.

\item The squeezing and sensitivity analysis conducted thus far in this paper is preliminary, as the coherent feedback network studied in this paper consists of a SISO quantum plant and a SISO  quantum controller. A comprehensive understanding of squeezing and sensitivity analysis and their applications in the design of quantum coherent feedback networks is one of our major future research goals.

\item Due to  the pole-zero duality explored in the paper, a linear quantum system must be either unstable or non-minimum phase, or even both.  It is well-known in classical linear systems theory that   unstable poles and unstable zeros impose  fundamental performance limitation in controller design,  thus the pole-zero duality of linear quantum systems will naturally lead to performance limitation in quantum controller design. Exploring this will be one of our future research directions.

\end{itemize}

\section*{Acknowledgment}
The authors thank the fruitful discussions with Professors Tongwen Chen, Xiang Chen and Long Wang. The authors thank the anonymous reviewers for detailed advice and constructive suggestions.


\bibliographystyle{IEEEtran}
\bibliography{gzhang.bib}

\end{document}